%% file: main.tex
\documentclass[sigconf]{acmart}
\usepackage{graphicx}
\usepackage{subcaption}
\usepackage{hyperref}
\usepackage{url}
\usepackage{enumitem}
\usepackage{booktabs}
\usepackage{balance}
\usepackage{stfloats}
\usepackage{pgfplots}
\pgfplotsset{compat=1.16}
\usepackage{xcolor}
\usepackage{tikz}
\usepackage{accents}

\AtBeginDocument{%
  \providecommand\BibTeX{{%
    \normalfont B\kern-0.5em{\scshape i\kern-0.25em b}\kern-0.8em\TeX}}}

\setcopyright{rightsretained}
\copyrightyear{2022}
\acmYear{2022}
\acmDOI{XXXXXXX.XXXXXXX}

\acmConference[Conference acronym 'XX]{Make sure to enter the correct
  conference title from your rights confirmation emai}{June 03--05,
  20XX}{Woodstock, NY}
\acmPrice{15.00}
\acmISBN{978-1-4503-XXXX-X/18/06}

\acmSubmissionID{1033}

\begin{document}

\title{Contrastive Pre-training for Deep Session Data Understanding}


\author{Zixuan Li}
\authornote{Zixuan Li is under the Industrial Postgraduate Program supported by Singapore Economic Development Board, Shopee Singapore Private Limited and National University of Singapore.}
\email{zixuan@comp.nus.edu.sg}
\affiliation{%
  \institution{Shopee Singapore Private Limited}
  \country{}
}
\affiliation{%
  \institution{National University of Singapore}
  \country{}
}

\author{Lizi Liao}
\email{lzliao@smu.edu.sg}
\affiliation{%
  \institution{School of Computing and Information Systems}
  \country{}
}
\affiliation{%
  \institution{Singapore Management University}
  \country{}
}

\author{Yunshan Ma}
\email{yunshan.ma@u.nus.edu}
\affiliation{%
  \institution{National University of Singapore}
  \country{}
}

\author{Tat-Seng Chua}
\email{dcscts@nus.edu.sg}
\affiliation{%
  \institution{National University of Singapore}
  \country{}
}

\renewcommand{\shortauthors}{Li, et al.}

\begin{abstract}
Session data has been widely used for understanding user's behavior in e-commerce. Researchers are trying to leverage session data for different tasks, such as purchase intention prediction, remaining length prediction, recommendation, \textit{etc}., as it provides context clues about the user's dynamic interests. However, online shopping session data is semi-structured and complex in nature, which contains both unstructured textual data about the products, search queries, and structured user action sequences. Most existing works focus on leveraging the coarse-grained item sequences for specific tasks, while largely ignore the fine-grained information from text and user action details. In this work, we delve into deep session data understanding via scrutinizing the various clues inside the rich information in user sessions. Specifically, we propose to pre-train a general-purpose User Behavior Model (UBM) over large-scale session data with rich details, such as product title, attributes and various kinds of user actions. A two-stage pre-training scheme is introduced to encourage the model to self-learn from various augmentations with contrastive learning objectives, which spans different granularity levels of session data. Then the well-trained session understanding model can be easily fine-tuned for various downstream tasks. Extensive experiments show that UBM better captures the complex intra-item semantic relations, inter-item connections and inter-interaction dependencies, leading to large performance gains as compared to the baselines on several downstream tasks. And it also demonstrates strong robustness when data is sparse. 
\end{abstract}





\maketitle

\input{intro}

\input{related}

\input{method}

\input{exp}

\section{Conclusion}
In this work, we proposed to pre-train a general purpose User Behavior Model via contrastive learning on large-scale e-commerce session data, which can be used for different downstream tasks. The two-level architecture of UBM allows the model to leverage both the textual information and complicated interaction clues. The proposed augmentation strategies together with the two-stage pre-training scheme further help the model to understand session data better. Extensive experiments on several downstream tasks verified the effectiveness of our proposed approach. We show that with the well-trained UBM model, the task-specific heads can be relatively simple and the learning process becomes more efficient.

Several works empirically suggest that using larger batch size leads to better performance for contrastive learning \cite{SimCLR}, while there are also researchers argue that it depends on the quality of the positive and negative pairs leveraged \cite{robinson2021contrastive, kalantidis2020hard}. In the future, we plan to consider improving the quality of negative samples for more advanced and robust session understanding results.


\bibliographystyle{ACM-Reference-Format}
\bibliography{reference}










\end{document}

%% file: intro.tex
\begin{figure}[!htb]
	\centering
	\vspace{+0.3cm}
	\includegraphics[width=0.99\linewidth]{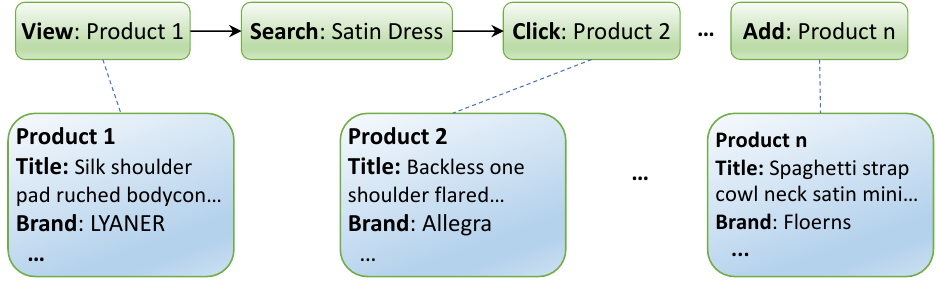}
	\vspace{-0.1cm}
	\caption{Illustration of a typical e-commerce user session. The user
interacts with products, search for products and add target item into cart through a sequence of interactions.}
	\vspace{-0.3cm}
	\label{example}
\end{figure}

\section{Introduction}
\label{sec:intro}
A session refers to a sequence of user interactions with an application interface that take place within a given time frame. It stores rich information about the user's activities on the 
application platform. For example, as the session illustrated in Figure \ref{example}, an online-shopping customer can issue a search query and perform various actions on the retrieved products, such as clicking, viewing, adding to cart, \textit{etc}. Such data can be very useful for understanding a user's needs so as to provide better services. Hence, we see various methods developed over session data for a wide variety of tasks \cite{srs_survey,AIR}, such as purchase intention prediction, remaining length prediction, item recommendation, \textit{etc}. 

Existing works mostly make use of the structured item sequences only, while largely ignore all the other information details such as text description or user actions. To be more specific, they take the sequence of the user interacted item IDs as input to understand user's behavior \cite{srs_survey, prod2bert}. Generally speaking, there are two groups of such models: (a) Markov-based models and (b) deep neural network (DNN)-based models. In the first group, the Markov-based models are developed early and show reasonably well performance in modelling sequence data. Hence, it has been widely applied in various scenarios \cite{rendle2010fpmc,m3pp}. However, it holds the assumption that the next state is only dependent on the current state, which makes it short-sighted and incapable in understanding the user's dynamically changing intention. In the second group, recurrent neural networks (RNNs) become more and more popular \cite{Diamantaras}. It allows to model sequential data and facilitates understanding the sequence as a whole, where the dynamically-changing user intention can be captured. At the same time, graph neural networks (GNNs) are also widely adapted for session modelling \cite{Wu_Tang_Zhu_Wang_Xie_Tan_2019}. Each item sequence is represented in a graph structure to incorporate the complex relations when generating user or item embeddings. Recently, with the proof of effectiveness of Transformers, many researchers try to model the sequential session data using Transformer-based models, such as BERT4Rec \cite{bert4rec} and Prod2BERT \cite{prod2bert}. Better performance results have been achieved.

However, e-commerce session data is generally semi-structured with rich details. It consists of unstructured textual data about products and search queries, as well as structured interaction sequences. Existing methods fail to properly model the data holistically or capture the complex relations. For example, not only the textual information about products and queries is under-explored, but also the effect of different actions (such as \textit{search, click,} \textit{add to cart,} \textit{etc.}) is largely ignored \cite{srs_survey}. Nonetheless, these textual information, including product titles and search queries, provides important clues about user preferences. Moreover, when understanding the overall context implied by the sequence, a \textit{click} action would has rather different effect comparing with an \textit{add to cart} action. As evidenced in research works such as \cite{AIR}, it is necessary to distinguish different actions in session modelling and user behavior understanding. Last but not least, with simple sequence of interacted item IDs only, models usually face data sparsity issue and cannot perform well when there is no enough interactions with new items \cite{ijcai2019p883}.


This motivates us to build a general and universal model to capture all these details and gain better session data understanding, which has the potential to systematically benefit all related downstream tasks. Nonetheless, there are several challenges. First of all, instead of modelling sessions as item ID sequences like in BERT4Rec \cite{bert4rec} and Prod2BERT \cite{prod2bert}, incorporating unstructured textual information introduces difficulties in both processing and modelling parts. The text sequence for each item and the item sequence are naturally in different granularity levels. Secondly, to ensure decent session understanding performance, large neural networks usually require massive labeled data to train from scratch. 
Although the current popular `pre-train then fine-tune' scheme signals a viable way \cite{SimonyanZ14a,bert,lan2019albert,yao2019kg}, those large-scale pre-trained language models require well-designed training objectives in order to learn the proper knowledge \cite{pre-train_survey}.

Fortunately, contrastive learning \cite{hadsell2006dimensionality} has brought prosperity to numbers of machine learning tasks by being a powerful unsupervised representation learning approach. Substantial performance gains have been reported in computer vision  \cite{he2020momentum,SimCLR} and natural language processing \cite{yan2021consert,simcse}. They believe that good representation should be able to identify semantically close neighbors while distinguish non-neighbors. Intuitively, in e-commerce user interaction data, sessions of similar user preference and background should resemble each other, while sessions with different user interests should be far apart from each other. Similarly, at the item level, similar items or high related items should be close to each other. Hence, by augmenting similar data samples to form positive pairs and training on large-scale session data with self-learning, it is possible to scrutinize the subtle clues inside sessions to better understand user behaviors \cite{technologies9010002}.

In this work, we thus propose to pre-train an universal general-purpose User Behavior Model (UBM) for semi-structured e-commerce session data, which can be easily leveraged by different downstream tasks. It models the heterogeneous data via a two-level Transformer-based architecture and captures the various relations via self-supervised contrastive learning with sophistically designed data augmentation strategies. Specifically, in the two-level network structure, the first level Transformer block focuses on learning the textual information of an item or a query together with the specific user action, which is realized by organizing the special action token and product related text or query as an input sequence. The second level Transformer block targets at capturing the meaning behind the sequence of interactions, which is realized by taking the sequence of (\textit{action, item/query}) tuples as input. It works in the higher level to infer the relations among different user interactions. To facilitate proper learning of the two-level architecture, we deliberately design a two-stage pre-training scheme with item-level and session-level augmentation strategies to allow the model to self-learn the heterogeneous relations lying in different granularities. 

The main contributions of our work are summarized below:
\begin{enumerate} 
\item We take the step towards AI efficiency revolution for the e-commerce industry, where one universal model can work for different purposes.
\item We propose to pre-train on e-commerce session data with a well-designed two-level Transformer-based architecture named UBM, which handles semi-structured data with actions and other rich text details seamlessly. 
\item We train the general-purpose UBM with two stages of self-learning objectives to gain better understanding on user behavior. Different session data augmentation strategies are innovatively designed to facilitate the learning process.
\item Extensive experiments on real-world e-commerce dataset demonstrate the effectiveness and applicability of our proposed UBM model, which can be easily adapted to support various downstream tasks with good performance. 
\end{enumerate}

%% file: related.tex
\section{Related Work}
\label{sec:related}

\subsection{E-commerce Session Data Modelling}
E-commerce session data is commonly used by different tasks, as it carries essential information about a user's preference. From the application perspective, it is widely used for recommendation task \cite{rendle2010fpmc,Hidasi2016SessionbasedRW} and other tasks such as purchase intention prediction \cite{m3pp,Diamantaras}, remaining session length prediction \cite{Gupta2020RLP}, \textit{etc}. 

In the most popular sequential recommendation task, methods can be categorized into Markov-based models \cite{mdp2002,rendle2010fpmc,m3pp} and DNN-based models \cite{Hidasi2016SessionbasedRW,Diamantaras,Gupta2020RLP,SakarPIP}. For example, FPMC brings matrix factorization and Markov chains together to capture the transition in a user behavior sequence so as to do next basket recommendation \cite{rendle2010fpmc}. However, Markov-based models is not able to capture long term relations, which later has been address by RNNs. Therefore, as one of the pioneer work, \citet{Hidasi2016SessionbasedRW} introduced GRU4Rec which models the whole session using RNN for session-based recommendation. Other than RNN, there are also attempts on using Convolutional Neural Network (CNN) and GNN. For example, \citet{CNN2018} showed that CNN performs better than RNN when the data does not contain strong sequential signals. At the same time, SR-GNN modelled session data into graph structure to capture complex transitions among items \cite{Wu_Tang_Zhu_Wang_Xie_Tan_2019}. Nevertheless, GNN-based approaches may suffer from popularity bias. As suggested by \cite{niser}, this bias is related to the norm of the learned vectors. Hence, the authors tried to address it by normalizing the representations. Very recently, many works show that Transformer is rather powerful in session modelling. For instance, Bert4Rec \cite{bert4rec} and Prod2BERT \cite{prod2bert} leveraged Transformer and achieved better performance results. At the same time, they also face the downside of using interacted item ID sequences, whereby their performance can be easily affected by data sparsity issue and being incapable in handling new items \cite{ijcai2019p883}.

Other than recommendation, session data is also used by other important tasks. \citet{m3pp} proposed a Markov modulated marked point process model to do early detection of user exit via the session data. It manages to leverage the time duration on each page to better understand the user's intention. In \cite{Gupta2020RLP}, they managed to do remaining session length prediction with the hypothesis that session lengths are Weibull distributed. It models the parameters via a RNN over the feature-based session data.

However, most of these usage of session data only focus on the structured item interaction sequences while largely ignored the textual information or the effect of different actions \cite{srs_survey}. Some works tried to leverage product attributes also \cite{lta}, but the usage of the textual information needs proper selection beforehand. At the same time, a few works started to emphasize the importance of identifying different actions \cite{AIR}, but it still remains under-explored.
In addition, to the best of our knowledge, there is a lack of efforts to tackle the session data modelling problem by leveraging both textual information and interaction sequences for general-purpose user behavior understanding.

\subsection{Pre-training}
Pre-training is found to be effective in different research areas, such as computer vision \cite{SimonyanZ14a,resnet} and natural language processing \cite{elmo,bert,gpt-3}. It helps to generalize representation learning and acts as a regularizer to avoid overfitting issue on downstream tasks with small dataset \cite{Dumitru_why}. BERT is a one of the most successful language models. It is pre-trained with a masked language modeling objective and allows bi-directional representation learning by conditioning on both left and right context of the sentence via Transformer \cite{bert}. Many variants of it further improved the performance \cite{roberta,clark2020electra}. For example, ELECTRA does replaced token detection instead of the masked token prediction, which is more efficient. The latest version of GPT model also show its powerfulness on language generation after being pre-trained on huge amount of data \cite{gpt-3}. Vision-language pre-trained models such as LXMERT \cite{lxmert} pre-train on both image and text for universal cross-modal representations. 
There also exist various table pre-training frameworks to assist downstream tasks \cite{dong2022table}. However, pre-training user session data is a relatively under-explored problem. \citet{prod2bert} use item interaction sequence as the input, and propose to pre-train a BERT-like model (Prod2BERT) for this kind of item sequence based user session data. However, item sequences and human language are fundamentally different. The vocabulary size would go extremely large when there are huge amount of unique items.

\subsection{Contrastive Learning}
Contrastive learning aims to learn through the objective of embedding augmented versions of the same sample close and pushing unrelated or negative samples away \cite{technologies9010002}. It has gain lots of attention in different research areas, as it is a simple and effective approach. In order to learn properly, the key of contrastive learning is to choose the appropriate augmentation strategies for generating positive and negative pairs \cite{goodview}. 
In computer vision, researchers usually augment the image data for constructing positive pairs via random cropping and resizing, horizontal flipping, Gaussian blurring, and so on \cite{SimCLR}. Similarly, in natural language processing community, some common augmentation methods  \cite{yan-etal-2021-consert} are word deleting, word shuffling, back translation, \textit{etc}. Different with these commonly used approaches, \citet{simcse} introduced SimCSE with a simple but effective augmentation method, where the positive sentence pairs are constructed by passing through the model with different dropout masks. Furthermore, \citet{Zhu2021ContrastiveLO} used three data augmentation strategies to do contrastive learning on structured query-document pair sequences for document ranking. Other than broadly augmenting positive samples, there are also works focusing on generating hard negatives while remaining self-supervised \cite{robinson2021contrastive, kalantidis2020hard}. 

However, there exists no augmentation strategy specifically tailored for session data so far. And to understand the contrastive learning, \citet{uniform_align} introduced the concept of uniformity and alignment, and it provides a good way to verify whether the contrastive objective optimizes these properties. Hence, it is later being widely used by different contrastive learning works as a measure for explaining the models.


%% file: method.tex
\section{Approach}
\label{sec:method}

\begin{figure*}[!htb]
\centering
\vspace{-0.2cm}
\includegraphics[width=1\textwidth]{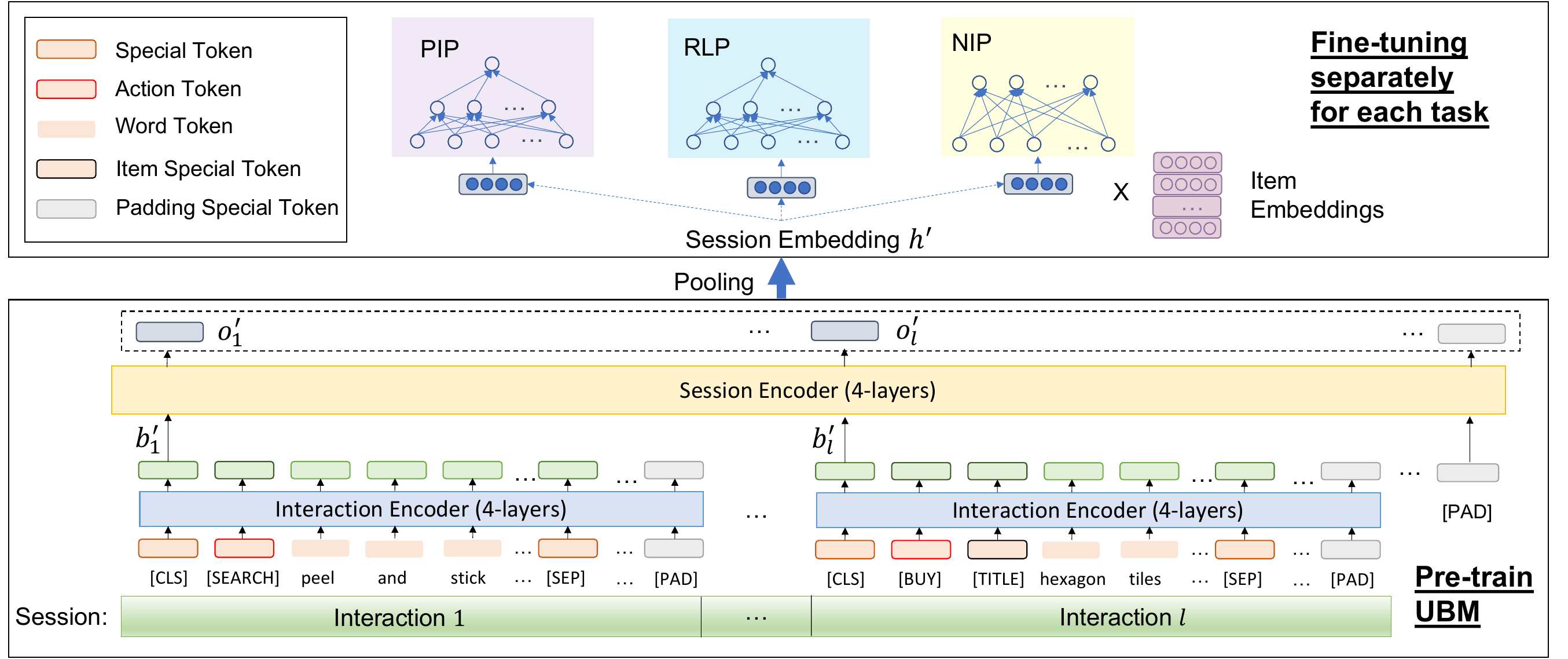}
\caption{Overview of the proposed \textbf{UBM} model. It is designed to be a two-level hierarchical structure with two-stage of pre-training. The low-level BERT-based Interaction Encoder is firstly pre-trained to captures intra-item semantics and inter-item connections. Then the whole UBM model is further pre-trained to encourage the high-level Transformer-based Session Encoder to learn inter-interaction dependencies, and allow the Interaction Encoder to further capture inter-action relations. After pre-training, a few simple task-specific layers are plugged in for downstream task fine-tuning.}
\vspace{-0.3cm}
\label{fig:UBM}
\end{figure*}

The overall architecture of our approach is illustrated in Figure \ref{fig:UBM}. It contains two major parts: pre-training UBM and fine-tuning for down-stream tasks. For pre-training the universal UBM model on session data, we design a generic hierarchical Transformer-based network structure. The low-level Interaction Encoder captures intra-item semantics and inter-item connections, and the high-level Session Encoder learns inter-interaction dependencies. The learning is enabled via a two-stage pre-training scheme with an innovative set of augmentation strategies. For fine-tuning, we simply add a few task-specific layers to benefit from the well-learned starting point.

For ease of illustration, we first briefly introduce the session data here. An e-commerce session \(s\) consists of a sequence of ordered user \textit{interactions} happened during the session, \textit{i.e}. \(s=[b_1, b_2, \cdots, b_l]\) where \(b\) refers to a specific \textit{interaction} and \(l\) is the length of the session. Each \textit{interaction} can be described by an action \(a\), together with a product item \(p\) or a text query \(q\), that is \(b=(a,p)\) or \(b=(a,q)\) respectively. As a result, each session is represented as a sequence of action-text pairs or say \textit{interactions} which are used as the data input in the UBM model.

\subsection{UBM Network Architecture}
The UBM model has a two-level hierarchical architecture formed by Transformer blocks. It is designed to encode both textual data and interaction sequences, so as to capture the heterogeneous relations and semantics in different granularities.

\subsubsection{Input Session Data Details}
Each session $s$ is treated as a long sequence. Inside which, there are four key actions that we consider: \textit{search, view, add to cart} and \textit{buy}.  They are represented with four special action tokens in the input sequence: `[SEARCH]', `[VIEW]', `[ADD]' and `[BUY]'. As almost all item click actions are followed by a product view, we use `[VIEW]' as a more general form that covers item click action here.
As for the product or the search query, they can be expressed in terms of text. We formulate the item text to be a concatenation of product title, product category and product attributes. To differentiate the different elements forming the item text, we put special tokens before the start of each element, which are `[TITLE]', `[CATEGORY]' and the specific attribute names. For example, if the attribute name is `size' and the attribute value is `large', it is formatted as `[SIZE] large'. 

\subsubsection{Low-level Interaction Encoder}
The first level of the UBM is instantiated with a 4-layer BERT \cite{bert} which consists of embedding layers and Transformer blocks. It works as an Interaction Encoder for encoding the sequence of text tokens describing an \textit{interaction}. As mentioned in section 3.1, other than special tokens like `[CLS]', `[SEP]', \textit{etc.}, there are some new special tokens involved, such as the special action tokens and the item special tokens. They are added to the vocabulary for the Interaction Encoder to learn more precisely. After the token sequence going through the Interaction Encoder, the embedding of the `[CLS]' token is taken as the \textit{interaction}'s embedding \(b'\), and then feed into the Session Encoder as input.

\subsubsection{High-level Session Encoder}
For the second level of the UBM model, we use another 4-layer Transformer block as the Session Encoder to encode the ordered sequence of \textit{interaction} embeddings \([b_1', b_2', \cdots, b_l']\) generated by the first level. It enables the model to learn the overall user behavior of the session. As the sequence order affects the user behavior understanding, we add the each vector in \([b_1', b_2', \cdots, b_l']\) with its corresponding sinusoidal positional embeddings before feeding them into the Transformer. Then, the output embeddings \([o_1', o_2', \cdots, o_l']\) of the \textit{interactions} after going through the Transformer block are then averaged to generate the final session representation \(h'\).

\subsection{Two-stage Pre-training}
The pre-training consists of two stages of contrastive learning. At the first stage, Interaction Encoder is pre-trained. We adapt the pre-trained BERT to our data for better encoding item information. At the second stage, the whole model is pre-trained on session data. 

\subsubsection{Pre-training Stage 1}
As the Interaction Encoder is BERT-based, we further adapt it to the e-commerce domain by doing contrastive learning on item text. To do self-supervised contrastive pre-training for the Interaction Encoder, we firstly do data augmentation at item level. Two augmentation strategies are used, i.e. \textit{Item Token Masking} (Figure \ref{fig:itm}) and \textit{Next Item Pairing} (Figure \ref{fig:nitm}). The item-next\_item pair is collected from the pre-training session dataset, and they automatically form a positive pair \(p\) and \(p^{++}\) in a batch with batch size set to N. The 2N items in the N item-next\_item pairs are applied with \textit{Item Token Masking} to obtain  \(p^+\). After going through the Interaction Encoder, we obtain the corresponding hidden vectors for the item data respectively: \(b'\), \(b'^+\) and \(b'^{++}\). Mathematically, the stage 1 pre-training is with the following two losses added:
\begin{align}
\ell_1=-\sum_{i=1}^{2N} log\frac{e^{sim(b^\prime_i,b_i^{\prime+})/\uptau}}{\sum_{j\ne i}e^{sim(b^\prime_i,b_j^{\prime+})/\uptau}},
\end{align}
\vspace{-0.2cm}
\begin{align}
\ell_2=-\sum_{k=1}^N log\frac{e^{sim(b^\prime_k,b_k^{\prime++})/\uptau}}{\sum_{q\ne k}e^{sim(b^\prime_k,b_q^{\prime++})/\uptau}},
\end{align}
Here $sim(\cdot)$ is a cosine similarity function and $\uptau$ is the temperature hyper-parameter. Within the batch, $h'_j$ is treated as a negative example of $h'_i$ when $i\neq j$, and $h'_q$ is treated as a negative example of $h'_k$ when $k\neq q$.

\subsubsection{Pre-training Stage 2}
\begin{figure}[!htbp]
     \centering
     \vspace{-0.1cm}
     \begin{subfigure}{0.4\textwidth}
         \centering
         \includegraphics[width=\textwidth]{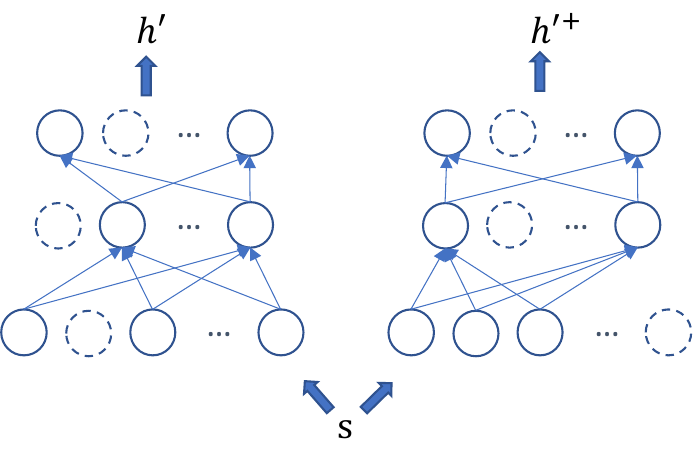}
         \caption{Dropout Masking}
         \label{fig:dm}
         \vspace{0.3cm}
     \end{subfigure}
     \begin{subfigure}{0.4\textwidth}
         \centering
         \includegraphics[width=\textwidth]{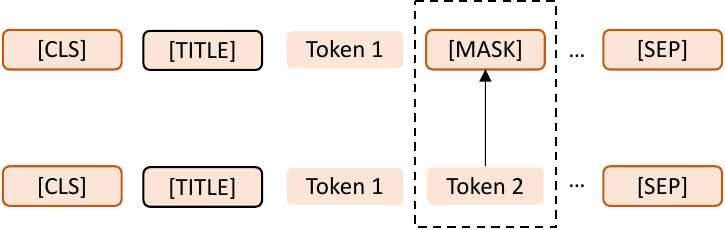}
         \caption{Item Token Masking}
         \label{fig:itm}
         \vspace{0.3cm}
     \end{subfigure}
     \begin{subfigure}{0.4\textwidth}
         \centering
         \includegraphics[width=\textwidth]{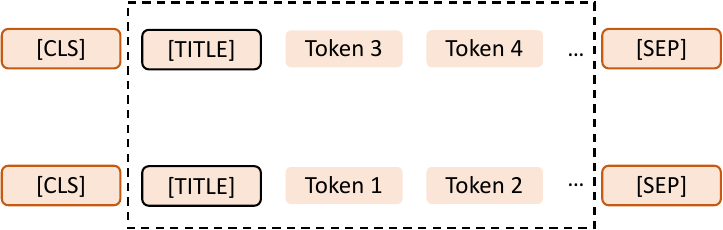}
         \caption{Next Item Pairing}
         \label{fig:nitm}
         \vspace{0.3cm}
     \end{subfigure}
     \begin{subfigure}{0.4\textwidth}
         \centering
         \includegraphics[width=\textwidth]{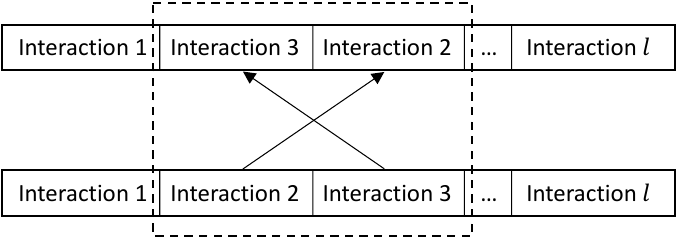}
         \caption{Behavior Reordering}
         \label{fig:br}
         \vspace{0.3cm}
     \end{subfigure}
     \begin{subfigure}{0.4\textwidth}
         \centering
         \includegraphics[width=\textwidth]{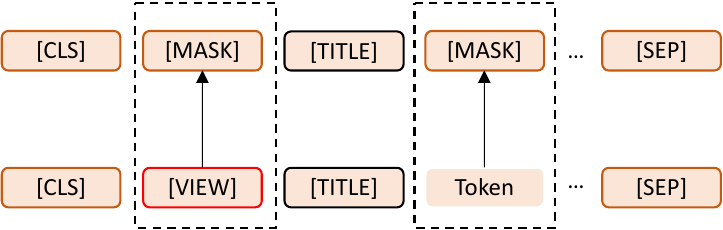}
         \caption{Action and Item Token Masking}
         \label{fig:aitm}
         \vspace{0.3cm}
     \end{subfigure}
        \caption{Augmentation Strategies. \textit{Dropout Masking} is automatically applied on all inputs; \textit{Item Token Masking} and \textit{Next Item Pairing} are used for item level augmentation; \textit{Behavior Reordering} and \textit{Action and Item Token Masking} are used for session level augmentation. By using these augmentation strategies in a two-stage pre-training, we manage to capture the complex intra-item semantic relations, inter-item connections and inter-interaction dependencies.}
        \label{fig:aug_str}
\end{figure}

In stage 2, we pre-train the UBM as a whole on session data with contrastive learning. To do self-supervised contrastive pre-training for the UBM, each session sample \(s\) is prepared with two augmented sessions \(s^+\) and \(s^{++}\). The two augmentation strategies applied are \textit{Behavior Reordering} (Figure \ref{fig:br}) and \textit{Action and Item Token Masking} (Figure \ref{fig:aitm}) respectively. After going through the model, we obtain the corresponding hidden vectors for the session data respectively: \(h'\), \(h'^+\) and \(h'^{++}\). Mathematically, the UBM is pre-trained with these two losses added:
\begin{align}
\ell_3=-\sum_{i=1}^N log\frac{e^{sim(h^\prime_i,h_i^{\prime+})/\uptau}}{\sum_{j\ne i}e^{sim(h^\prime_i,h_j^{\prime+})/\uptau}},
\end{align}
\begin{align}
\ell_4=-\sum_{i=1}^N log\frac{e^{sim(h^\prime_i,h_i^{\prime++})/\uptau}}{\sum_{j\ne i}e^{sim(h^\prime_i,h_j^{\prime++})/\uptau}},
\end{align}
for a batch with \(N\) session samples. Here $sim(\cdot)$ is a cosine similarity function and $\uptau$ is the temperature hyper-parameter. Within the batch, $h'_j$ is treated as a negative example of $h'_i$ when $i\neq j$.

\vspace{-0.1cm}

\subsection{Data Augmentation Strategies}
We further explain the different data augmentation strategies below, as it is the key for effective contrastive learning. We explore five augmentation strategies together to address the learning of different relations from the session data. An illustration of each strategy can be found in Figure \ref{fig:aug_str}.

\subsubsection{Dropout Masking} 
Dropout masking is found to be a simple and effective data augmentation method for text sequences \cite{simcse}. It is automatically applied to all inputs when we set dropout rate to non-zero. The dropout layers are the standard ones used in Transformer, and no extra dropout layers added.

\subsubsection{Item Token Masking}
BERT masked 15\% of the sentence tokens for masked language modelling, which obtained good results. We follow the setting to generate augmented data by applying random masking on the word tokens. Same as most research works, 10\% of them are replaced with `[MASK]' tokens, 5\% are replaced with random tokens, and 5\% are kept unchanged. 

\subsubsection{Next Item Pairing}
Given an item, we use its next item in the session sequence to form a positive pair. This augmentation strategy helps to learn item context information. As the first introducing this augmentation strategy, it is an innovative and effective try specifically tailored for session data.

\subsubsection{Interaction Reordering}
User's behavior sequence is highly variable and the overall interaction order matters. However, within a short distance, the order of the behavior should not affect the overall session understanding much. Therefore, we randomly choose two interactions from the session sequence and exchange their positions for data augmentation. This helps the UBM to be more robust and learn the true user intention behind each session. 

\subsubsection{Action and Item Token Masking}
This is similar to \textit{Item Token Masking}. As the \textit{interaction} data consists of both action token and product (or search query) tokens, to enable the model to learn the effect of different actions, the 15\% of token masking is applied on both special action tokens and item (or search query) text tokens.

\subsection{Task-specific Fine-tuning} 
\label{tasks}
With the UBM pre-trained on large-scale session data, different downstream tasks can then leverage the well-trained model as the encoder part to do task-specific fine-tuning. Serving as a general user behavior understanding model, only a few task-specific layers and a relatively small dataset are needed to obtain a good performance for the downstream tasks. 
Specifically, we choose three well-known tasks: Purchase Intention Prediction (PIP), Remaining Length Prediction (RLP) and Next Item Prediction (NIP) to verify our proposed approach. By knowing the answer for these three prediction tasks, the e-commerce platform can then make informed decisions based on whether the user is going to make a purchase, how long the session will still last and what is the next item he/she is going to interact with.

\subsubsection{Purchase Intention Prediction} 
PIP is a common task in e-commerce to predict whether a user is going to make a purchase. It would be very useful if the platform knows the answer in real time, so as to take corresponding actions, such as providing incentives, to improve the conversion rate. We append two feed-forward layers and one linear layer followed by a Sigmoid layer for fine-tuning the PIP task. It is trained with the Binary Cross Entropy Loss.
\begin{align}
\ell^{PIP}_{i}=-y_i log(\hat{y_i})+(1-y_i)log(1-\hat{y_i}),
\end{align}
where \(\hat{y}\) refers to the prediction, \(y\) denotes the ground truth and \(i\) refers to the \(i\)-th session.

\subsubsection{Remaining Length Prediction} 
By knowing how long the session will last, the platform can take corresponding actions timely. We append two feed-forward layers to output a number. It is then trained with the Mean Square Error Loss.
\begin{align}
\ell^{RLP}=\frac{\sum_{i=1}^{K}(\hat{y_i}-y_{i})^2}{K},
\end{align}
where \(K\) is the total number of training data, \(\hat{y}\) is the predicted length value and \(y\) is the ground truth length.

\subsubsection{Next Item Prediction}
NIP is quite popular recently. By knowing the next item that the user is going to interact with, the platform can take immediate actions accordingly. For the \(m\) products in the candidate pool, we generate the items' embeddings using the pre-trained UBM by treating each product as a session with no action tokens and with length \(l=1\):
\[v_i=UBM([p_i])\]
where \(i\) refers to the \(i\)th product and \(v_i\) refers to the corresponding item embedding. We use the candidate pool of items' embeddings \(V=\{v_1, v_2, ..., v_m\}\) to multiply with an session embedding \(h'\) to get the predicted score \(\hat{z}\) for each candidate item.
\[\hat{z_i}=h'V_i^T\]
A Softmax layer is then applied to get the probability \(\hat{y}\) over the candidate pool to predict the one to be interacted next.
\[\hat{y}=Softmax(\hat{z})\]
 The Cross Entropy Loss is used and can be written as the following:
\begin{align}
\ell^{NIP}_i=-\sum_{c=1}^{m}y_{i,c}log(\hat{y}_{i,c}),
\end{align}
where \(y\) is the binary indicator vector telling  whether the specific item is the ground truth of the given session.

%% file: exp.tex
\section{Experiment}
\label{sec:exp}

\subsection{Experiment Setup}
\subsubsection{Dataset} 
The dataset is provided by an e-commerce company spanning multiple countries. We randomly sample sessions with purchase from a given month. There are 2,949,868 sessions in total used for pre-training after some pre-processing. Generally speaking, for each purchase in a session, we back-trace up to 32 interactions until a previous purchase encountered or no more interactions found in the session. 
We limit each constructed session with length less than 32, because the majorities of session lengths are less than 32, and this enables efficient usage of GPU memory by avoiding too many padding tokens. With similar reasons, only the first 32 tokens of the product title is considered, and the overall length of the product text is limited to 64 tokens. 

As for downstream tasks, we randomly sample from all session data including both `sessions with purchase' and `sessions without purchase'. Training data is extracted from the log of a day following the month for pre-training, and validation and testing data are from the log of the following day. For PIP, as the data is imbalanced, over-sampling is done on those with purchase. For RLP, we randomly cut the session at a position before the 32nd interaction and prepare the label correspondingly. In both PIP and RLP, we ignored the input sessions shorter than 5. For NIP, we follow most existing works \cite{Wu_Tang_Zhu_Wang_Xie_Tan_2019,stamp} by ignoring items appeared less than 5 times and removed the item from validation/test set if it does not appear in the training set. Given a session \(s=[b_1, b_2, ..., b_l]\), we generate sequence and label pairs as \(([b_1],b_2)\), \(([b_1, b_2], b_3)\), \textit{etc}. To sum up, 66,120 training data and 10,424 validation/test data are prepared for the PIP task, 37,715 training data and 10,424 validation/test data for the RLP task, and 182,678 training data and 49,891 validation/test data for the NIP task. There are 27,860 unique products involved for NIP, and all are used in the NIP candidates pool.

\subsubsection{Evaluation Tasks and Metrics}  
As mentioned in Section \ref{tasks}, PIP, RLP and NIP are used as down-steam tasks for evaluation. We explain the metrics used for each of them below.

{\bf Purchase Intention Prediction.}
We use accuracy, Area Under Receiver Operating Characteristic (AUROC), F1 score and Cohen's Kappa (CK) as our metrics here. As a highly imbalanced binary classification problem, accuracy might not be a good choice. However, as both classes are important sometimes, we keep it as a rough reference. AUROC is a commonly used metric for imbalanced data, where AUROC=1 refers to perfect prediction. F1 Score combines both precision and recall into one metric, and it is a good indicator if one cares more about the positive class. CK is used to measure the agreement between two raters. We use it here to assess the model's performance. 

{\bf Remaining Length Prediction.}
As a regression problem, we use the standard Mean Absolute Error (MAE), Mean Squared Error (MSE) and Mean Squared Log Error (MSLE) as our metrics. They range from 0 to infinity and the smaller the better. In addition, the R\textsuperscript{2} score measures how well the regression predictions approximate the real data points. When R\textsuperscript{2}=1, it indicates a good fit.

{\bf Next Item Prediction.}
We evaluate the NIP performance with hit ratio and Mean Reciprocal Rank (MRR). The hit@K measures how often the correct item appears in the top K predicted products, while MRR@K cares more about the rank of the correct item in the top K predictions. MRR is calculated as 1 divided by the rank position. If it does not appear in the top K predictions, it is set to 0.

\subsubsection{Baselines} 
To verify the effectiveness of the proposed approach, we compare it with three groups of baselines as described below. We would like to verify the effectiveness of leveraging both textual information and interaction sequence by comparing UBM with the followings: general-domain pre-trained language models which get pre-trained on general textual data; domain adapted language models which get continue pre-trained on interaction data and e-commerce pre-trained session model which only uses item ID sequence. In addition, we also have the task specific methods to show the necessity of pre-training.

{\bf Task Specific Models.} 
For PIP, $RAND$ refers to generating random binary predictions. For $FIXED$ under RLP, we used the average remaining length of training data as prediction results. For NIP, we have $POP$ and \textit{SR-GNN} \cite{Wu_Tang_Zhu_Wang_Xie_Tan_2019}. $POP$ means that we use the most popular items appeared in training data as prediction results. \textit{SR-GNN} is one of the state-of-the-art method for session-based recommendation. There exists other next item recommendation models, however, most of them are not for within session sequence.

{\bf General-domain Pre-trained Language Models.}
We use pre-trained $BERT_{mini}$ \cite{bhargava2021generalization,DBLP:journals/corr/abs-1908-08962} (4 layers with hidden size 256), pre-trained $BERT_{8layer}$ (8 layers with hidden size 256) and also pre-trained $ELECTRA_{small}$ \cite{clark2020electra} (12 layers with hidden size 256) to generate embeddings for each interaction, and take the average over the interactions of the session to get the session embedding. Similar to ours, they are concatenated with the task specific layers for overall fine-tuning in the three tasks for evaluation. The three general-domain language models help to provide baselines on only using general textual data information.

{\bf E-commerce Pre-trained Model.} 
We continue pre-train $BERT_{mini}$ and $BERT_{8layer}$ on our pre-training dataset by letting them seeing all interactions. This is to provide a more fair comparision among the pre-trained models.
In addition, we also have $Prod2BERT$ \cite{prod2bert} as e-commerce pre-trained session model which treats each item as an ID token and each interaction sequence as a `sentence'. The authors suggest to pre-train a BERT-like model to learn session embeddings via masked `language' modelling task. We pre-train it using the same pre-training dataset. To make it comparable, we configure the number of its transformer layers to be 8 and embedding size to be 256, which is the same as UBM. As for `vocabulary' size, we set it to 1,500,000 which is the largest the available resource can handle. The most popular 1,500,000 items are used to construct the `vocabulary'. This is one of the shortage for $Prod2BERT$, where our method can handle unlimited number of new items. We then generate session embeddings using the pre-trained $Prod2BERT$ for PIP and RLP fine-tuning on the task-specific layers. As for NIP, we found that the results are near to zero, hence it is omitted from the results table. It might because nearly all items in the test set are not in the `vocabulary', which makes it not comparable. 



\vspace{-0.5cm}
\subsection{Implementation Details}
\label{implement}
\subsubsection{Pre-training Details} 
We use 4 Tesla V100 GPUs for pre-training. It takes around 8 hours for stage 1 pre-training with one epoch, and around 4 hours for stage 2. As for batch size, following the suggestion in \cite{SimCLR} that contrastive learning requires larger batch size, the effective batch size is set to 512 and 128 for stage 1 and stage 2 respectively, which is the maximum we can use given that the GPU memory is 32G. As for dropout masking, the dropout rate is set to 10\%. We set a scheduler for learning rate update, the first 10\% steps are for warm-up and the peak value is 3e-5 followed by a linear decrease till 0. The temperature hyper-parameter is set to 0.05.
As for both Interaction Encoder and Session Encoder, each transformer block is with 4 layers and the hidden size is set to 256. They are chosen in consideration with the available GPU memory, as each session data may consists of 32*64=2048 tokens.
\subsubsection{Fine-tuning Details} 
During fine-tuning, each task is trained on a single GPU with batch size set to 32. We used 3e-5 as the learning rate. All tasks are trained for 30 epochs. The whole structure including UBM and task-specific layers are fine-tuned together. And we use the model checkpoint with minimum loss on validation data for evaluating the test set.


\begin{table}[!htbp]
\small
\caption{\label{table:main}Results on PIP, RLP and NIP tasks. $\uparrow$ indicates the larger the better while $\downarrow$ indicates the smaller the better. The * symbol means that our performance improvement is statistically significant with p<0.05 against all baselines.}
\vspace{-0.2cm}
\renewcommand*{\arraystretch}{1.1}

\begin{tabular}{lllll}
\toprule
\multicolumn{5}{c}{Next Item Prediction}                                     \\ \hline
\multicolumn{1}{l|}{}                    & hit@10 $\uparrow$& MRR@10 $\uparrow$& hit@20 $\uparrow$& MRR@20 $\uparrow$\\ \hline
\multicolumn{1}{l|}{POP}     & 0.017 & 0.0072	& 0.0203 & 0.0074	  \\ 
\multicolumn{1}{l|}{SR-GNN}  & 0.0016	& 0.0005	& 0.0022	& 0.0006 \\ \hline
\multicolumn{1}{l|}{BERT$_{mini}$} & $\underline{0.4320}$	& $\underline{0.3011}$	& 0.4792	& $\underline{0.3044}$ \\ 
\multicolumn{1}{l|}{BERT$_{8layer}$}     & 0.4105	& 0.2728	& 0.4632	& 0.2764 \\ 
\multicolumn{1}{l|}{ELECTRA$_{small}$}     & 0.2556	& 0.1496	& 0.3037	& 0.1529 \\\hline
\multicolumn{1}{l|}{BERT$_{mini\_domain}$} & 0.4253	& 0.2763	& $\underline{0.4863}$	& 0.2805 \\ 
\multicolumn{1}{l|}{BERT$_{8layer\_domain}$} & 0.4189	& 0.2542	& 0.4833	& 0.2586 \\  \hline 
\multicolumn{1}{l|}{UBM$_{wo}$} & 0.3198	& 0.1562	& 0.4037	& 0.1619 \\ 
\multicolumn{1}{l|}{UBM$_{wo\_BERT}$} & 0.5351	& 0.3757	& 0.5945	& 0.3798 \\ 
\multicolumn{1}{l|}{UBM} & \bf0.5568$^\ast$ 	& \bf0.3975$^\ast$ 	& \bf0.6131$^\ast$ 	& \bf0.4013$^\ast$   \\ 
\bottomrule

\multicolumn{5}{c}{Purchase Intention Prediction}                                               \\ \hline
\multicolumn{1}{l|}{}                                   & Accuracy $\uparrow$& AUROC $\uparrow$ & F1 Score $\uparrow$& CK  $\uparrow$   \\ \hline
\multicolumn{1}{l|}{RAND}     & 0.4906	& 0.5045	& 0.2689	& -0.0107 \\ \hline
\multicolumn{1}{l|}{BERT$_{mini}$} & $\underline{0.6202}$	& 0.6636	&$\underline{0.3689}$	& $\underline{0.1601}$ \\ 
\multicolumn{1}{l|}{BERT$_{8layer}$}     & 0.4391	& 0.6662	& 0.3357	& 0.0800 \\ 
\multicolumn{1}{l|}{ELECTRA$_{small}$}    & 0.1754	& 0.4904	& 0.2985	& 0 \\ \hline
\multicolumn{1}{l|}{BERT$_{mini\_domain}$} & 0.4269	& $\underline{0.6697}$	& 0.3446	& 0.0891 \\ 
\multicolumn{1}{l|}{BERT$_{8layer\_domain}$} & 0.2085	& 0.5362	& 0.2942	& -0.0025 \\ 
\multicolumn{1}{l|}{Prod2BERT}  &0.6049	&0.6092	&0.3245	&0.1102\\ \hline
\multicolumn{1}{l|}{UBM$_{wo}$} &0.5042	&0.6861	&{0.3952}	&0.1649 \\ 
\multicolumn{1}{l|}{UBM$_{wo\_BERT}$} &0.6175	&0.7483	&0.4307	&0.2315 \\ 
\multicolumn{1}{l|}{UBM} & \bf0.7405$^\ast$ 	& \bf0.7606$^\ast$ 	& \bf0.4515$^\ast$ 	& \bf0.2960$^\ast$  \\  
\bottomrule

\multicolumn{5}{c}{Remaining Length Prediction}                                              \\ \hline
\multicolumn{1}{l|}{}                    & MAE $\downarrow$   & MSE  $\downarrow$    & MSLE $\downarrow$  & R$^2$ $\uparrow$                 \\ \hline
\multicolumn{1}{l|}{FIXED}     &8.7757	&\bf202.2618	&1.0518	&1.79E-07 \\ \hline
\multicolumn{1}{l|}{BERT$_{mini}$} &8.7790	&216.4974	&1.0037	&-0.0142 \\ 
\multicolumn{1}{l|}{BERT$_{8layer}$}    &8.6732	&208.6623	&$\underline{0.9845}$	&0.0225 \\ 
\multicolumn{1}{l|}{ELECTRA$_{small}$}    &$\underline{8.6666}$	&207.8466	&0.9852	&$\underline{0.0263}$ \\ \hline
\multicolumn{1}{l|}{BERT$_{mini\_domain}$} &\bf8.3665	&230.7037	&\bf0.8894	&-0.0808 \\ 
\multicolumn{1}{l|}{BERT$_{8layer\_domain}$} &8.8071	&210.284	&1.0047	&0.0149 \\ 
\multicolumn{1}{l|}{Prod2BERT}  &9.1908	&243.6677	&1.0140	&0.0134\\ \hline
\multicolumn{1}{l|}{UBM$_{wo}$} &8.7280	&208.6614	&0.9948	&0.0225 \\ 
\multicolumn{1}{l|}{UBM$_{wo\_BERT}$} &8.8139	&209.5886	&1.0097	&0.0182 \\ 
\multicolumn{1}{l|}{UBM} &8.6872	&\underline{207.7964}$^\ast$ 	&0.9855	&\bf0.0266$^\ast$  \\   
\bottomrule
\end{tabular}


\vspace{-0.6cm}
\end{table}

\subsection{Main Results}
As it shows in Table \ref{table:main}, with pre-trained UBM, we achieved much better performance across all downstream tasks. For example, in NIP, it outperforms \(BERT_{mini}\) for 28.9\% on hit@10 and 32\% on MRR@10. As for PIP, it achieves much better results comparing with the baselines, whereby the performance improvement is as much as 13.6\% on AUROC. In RLP, it also obtained generally good results across different metrics. This verifies the effectiveness and generality of our proposed UBM method.

\begin{figure}[t]
\centering
\vspace{+0.3cm}
\includegraphics[width=0.99\linewidth]{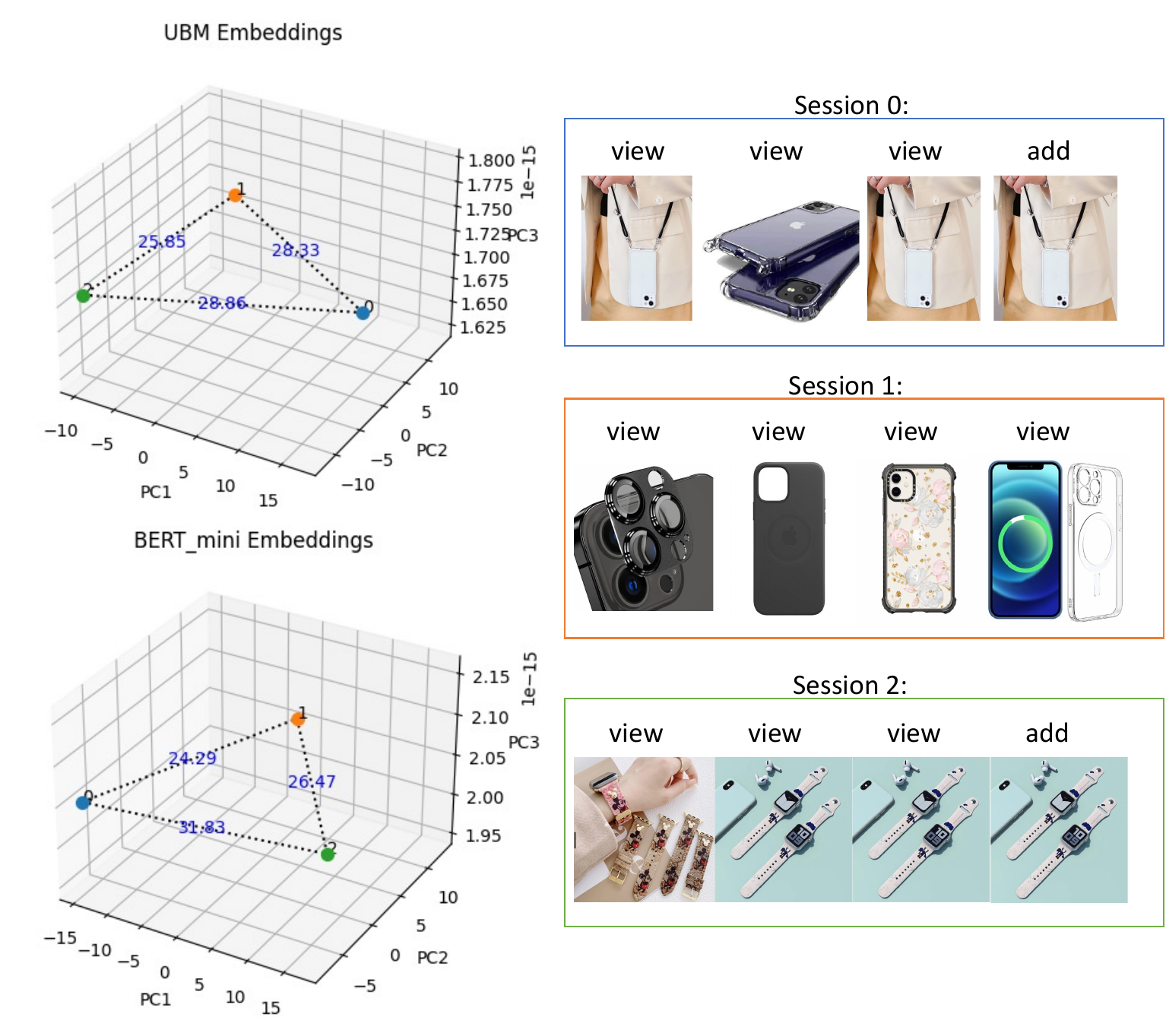}
\vspace{-0.1cm}
\caption{Visualization of example sessions' representations learned by UBM versus {BERT$_{mini}$}.}
\vspace{-0.5cm}
\label{concrete_example}
\end{figure}

When the UBM architecture is used but without pre-training (\(UBM_{wo}\)), it performs much worse which implies the importance of pre-training the UBM for general user behavior understanding. We also experiment with pre-training UBM without loading pre-trained BERT as starting point for Interaction Encoder (\(UBM_{wo\_BERT}\)), the performance is slightly worse. This shows the importance to start with a good pre-trained starting point.

As for the general-domain language models, \(BERT_{mini}\) has 4 layers, \(BERT_{8layer}\) has 8 layers and \(ELECTRA_{small}\) has 12 layers. Interestingly, the performance drops as the number of layers increase for both NIP and PIP. We suspect the NIP and PIP tasks are more sensitive to textual data, and the larger the model, the harder it can be fine-tuned.

It seems simply continue pre-training the language models does not help or even hurts the performance. This might suggest the fundamental difference between normal language sentences and the session interactions. Product titles are usually not a standard complete sentence in proper grammar, and we also concatenate it with short attribute phases and special tokens to form the \textit{interaction}. Hence, using the standard masked language modelling objective will not help to learn session data.

At the same time, we found that $Prod2BERT$ gives bad results on PIP and RLP. There are many possible reasons. From one hand, it cannot handle large number of unique items. One needs to limit the `vocabulary' size according to the available resources. This also shows the advantage of our method, as UBM can take care of any amount of new products as long as it can be properly represented by text information. On the other hand, it does not leverage the textual information, hence missing a lot of evidence for session understanding. Under NIP, the evaluation results of \textit{SR-GNN} are also not good. This might due to the fact that \textit{SR-GNN} does not see the pre-training data while only uses the task's training data. In addition, it only make use of the item ID sequences for learning. 

To visualize how the session embeddings are learned by different models which lead to the different performance, we find a difficult case: two user sessions browsing iPhone cases and one user session looking for iWatch strap. As shown in Figure \ref{concrete_example}, Session 0 and Session 2 represent similar user behavior where the user adds the item to cart after repetitive viewing, despite that Session 0 is buying iPhone case and Session 2 is buying iWatch strap. Session 1 refers to a continuous browsing of different variations of iPhone cases. UBM brings Session 0 and Session 2 closer and pushes Session 1 away from Session 0 than \(BERT_{mini}\) as it is trained to learn overall user behavior.

\begin{table}[!htbp]
\small
\caption{\label{table:ablation}Experiment results on pre-training UBM with one of the loss ignored.}
\vspace{-0.2cm}
\renewcommand*{\arraystretch}{1.1}

\begin{tabular}{lllll}
\toprule
\multicolumn{5}{c}{Next Item Prediction}                                     \\ \hline
\multicolumn{1}{l|}{}                    & hit@10 $\uparrow$& MRR@10 $\uparrow$& hit@20 $\uparrow$& MRR@20 $\uparrow$\\ \hline
\multicolumn{1}{l|}{UBM}      & 0.5568	& 0.3975	& 0.6131	& 0.4013  \\ \hline
\multicolumn{1}{l|}{UBM$_{no\_loss1}$}    & 0.5579	& 0.4022	& 0.6123	& 0.4060 \\
\multicolumn{1}{l|}{UBM$_{no\_loss2}$} & 0.5076	& 0.3678	& 0.5574	& 0.3713 \\ \hline
\multicolumn{1}{l|}{UBM$_{no\_loss3}$}     & 0.5487	& 0.3895	& 0.6047	& 0.3934 \\
\multicolumn{1}{l|}{UBM$_{no\_loss4}$} & 0.5504	& 0.3928	& 0.6053	& 0.3966 \\ \bottomrule

\multicolumn{5}{c}{Purchase Intention Prediction}                                               \\ \hline
\multicolumn{1}{l|}{}                                   & Accuracy $\uparrow$& AUROC $\uparrow$ & F1 Score $\uparrow$& CK  $\uparrow$   \\ \hline
\multicolumn{1}{l|}{UBM}      & 0.7405	& 0.7606	& 0.4515	& 0.2960  \\ \hline
\multicolumn{1}{l|}{UBM$_{no\_loss1}$}    & 0.7736	& 0.7212	& 0.3649	& 0.2271 \\
\multicolumn{1}{l|}{UBM$_{no\_loss2}$} & 0.7506	& 0.7412	& 0.4382	& 0.2868 \\ \hline
\multicolumn{1}{l|}{UBM$_{no\_loss3}$}     & 0.7494	& 0.7265	& 0.4247	& 0.2721 \\
\multicolumn{1}{l|}{UBM$_{no\_loss4}$} & 0.7497	& 0.7296	& 0.3916	& 0.2383 \\ \bottomrule

\multicolumn{5}{c}{Remaining Length Prediction}                                              \\ \hline
\multicolumn{1}{l|}{}                    & MAE $\downarrow$   & MSE  $\downarrow$    & MSLE $\downarrow$  & R$^2$ $\uparrow$                 \\ \hline
\multicolumn{1}{l|}{UBM}     & 8.6872	& 207.7964	& 0.9855	& 0.0266  \\ \hline
\multicolumn{1}{l|}{UBM$_{no\_loss1}$}   & 8.8344	& 209.5136	& 1.0129	& 0.0185 \\
\multicolumn{1}{l|}{UBM$_{no\_loss2}$} & 8.6709	& 208.4276	& 0.9840	& 0.0236 \\ \hline
\multicolumn{1}{l|}{UBM$_{no\_loss3}$}  & 8.8044	& 208.9460	& 1.0079	& 0.0212 \\
\multicolumn{1}{l|}{UBM$_{no\_loss4}$} & 8.8270	& 209.3905	& 1.0122	& 0.0191 \\ \bottomrule
\end{tabular}

\vspace{-0.2cm}
\end{table}

\subsection{Ablation Study}
To further understand how different augmentation strategies help on modelling session data, we pre-train UBM with one corresponding loss ignored at a time, and used them for evaluations. The rest of the implementation details are the same as described in section \ref{implement}. The results can be found in Table \ref{table:ablation}. For example, ${UBM_{no\_loss1}}$ refers to the UBM without learning from the \textit{Item Token Masking} augmentation, ${UBM_{no\_loss2}}$ refers to the UBM without learning from the \textit{Next Item Pairing} augmentation, ${UBM_{no\_loss3}}$ refers to the UBM without learning from the \textit{Behavior Reodering} augmentation, and ${UBM_{no\_loss4}}$ refers to the UBM without learning from the \textit{Action and Item token Masking} augmentation.

By removing loss2 which corresponds to \textit{Next Item Pairing} augmentation method, the NIP's performance drops a lot. This is intuitive to understand as this augmentation strategy is essential for NIP task. On the contrary, removing loss2 has the least effect on PIP. As for RLP, it seems rely more on loss1 and loss4, whereby both are token masking methods.

All performance drops when removing any of the losses, except removing loss1 for NIP. According to the experiment results, it seems NIP is not affected by loss1. However, loss1 is quite important for other tasks, and it did not harm NIP's performance. In short, each augmentation strategy has its own strengths on addressing different relations of the session data. By using all of them, the UBM gains generality on user behavior modelling.

\subsection{Robustness on Sparse Items}

\begin{figure}[!htbp]
\centering
\vspace{+0.3cm}
\includegraphics[width=0.99\linewidth]{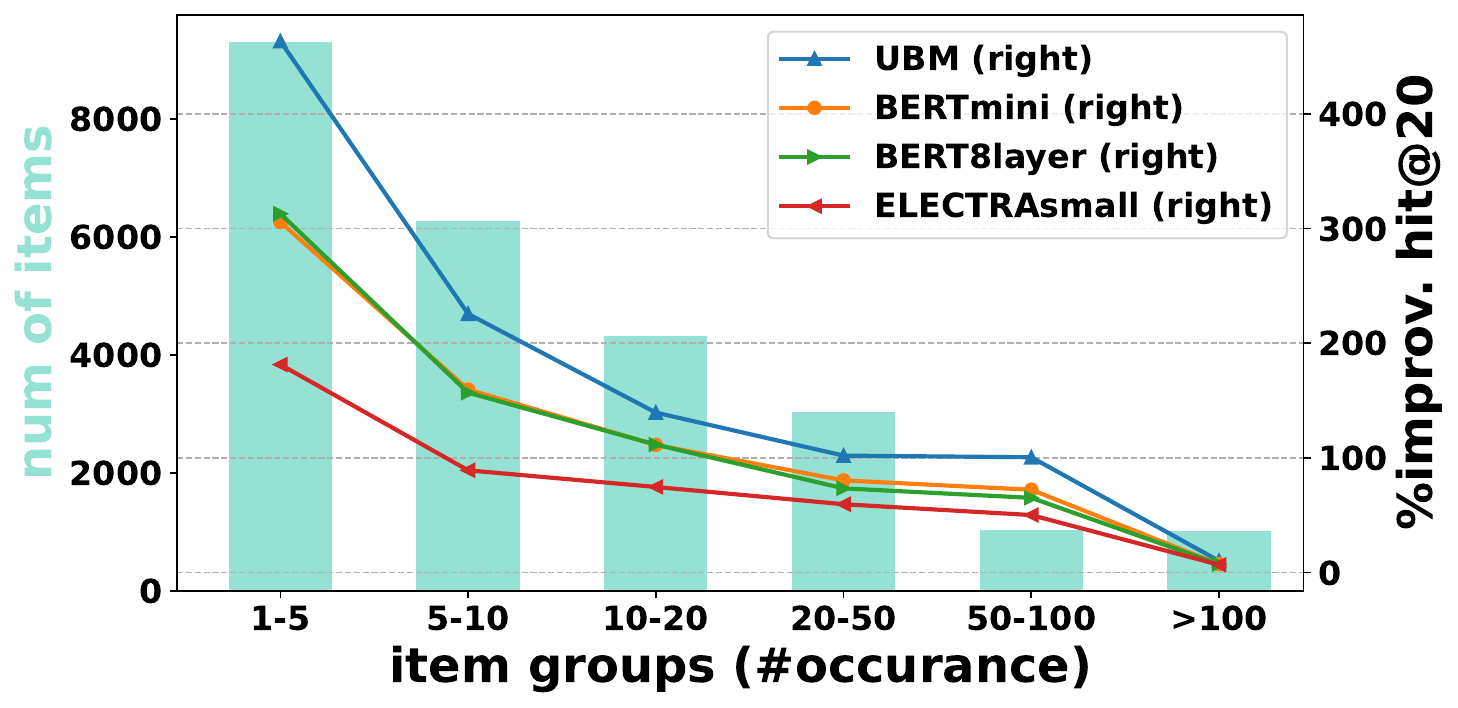}
\vspace{-0.1cm}
\caption{Models' NIP performance on different item groups.}
\label{sparsity}
\end{figure}

When a model makes use of item ID as input, it is generally limited by sparsity issue. Our UBM model and pre-trained language models like $BERT_{mini}$, $BERT_{8layer}$ and $ELECTRA_{small}$ avoids this problem by using text to represent item. To demonstrate our robustness on sparse data, we divide test dataset into different sparsity groups according to their occurrence in the training dataset, and then run NIP experiments on each data group. The results are draw in Figure \ref{sparsity}. The performance improvement is calculated with reference to \textit{SR-GNN} as a baseline. As it shows, our UBM model achieved much higher performance gain on sparse items. 

\subsection{Alignment and Uniformity} 
\begin{figure}[!htbp]
     \flushleft
     \begin{minipage}{0.22\textwidth}
      \begin{tikzpicture}
        \begin{axis}[
            title={NIP},
            xlabel=$\ell_{uniform}$,
            ylabel=$\ell_{align}$,
            width=4.2cm,
            height=4.2cm,
            legend style={at={(1.2,0.5)},anchor=west},
            axis equal
          ]
          \pgfplotstableread{data2NIP.txt}\mydatasNIP;
          \addplot+[only marks] table [x=uniformity,y=alignment]{\mydatasNIP};
          \pgfplotstableread{data3NIP.txt}\mydatatNIP;
          \addplot+[only marks] table [x=uniformity,y=alignment]{\mydatatNIP};
          \pgfplotstableread{data4NIP.txt}\mydatafNIP;
          \addplot+[only marks] table [x=uniformity,y=alignment]{\mydatafNIP};
          \pgfplotstableread{data5NIP.txt}\mydataffNIP;
          \addplot+[only marks] table [x=uniformity,y=alignment]{\mydataffNIP};
          \pgfplotstableread{data6NIP.txt}\mydatasNIP;
          \addplot+[only marks] table [x=uniformity,y=alignment]{\mydatasNIP};
         \end{axis}
         \end{tikzpicture}
         \end{minipage}
         \begin{minipage}{0.15\textwidth}
          \begin{tikzpicture}         
            \begin{axis}[
                title={PIP},
                xlabel=$\ell_{uniform}$,
                ylabel=$\ell_{align}$,
                width=4.2cm,
                height=4.2cm,
                legend style={at={(1.2,0.5)},anchor=west},
                axis equal
              ]
              \pgfplotstableread{data2PIP.txt}\mydatasPIP;
              \addplot+[only marks] table [x=uniformity,y=alignment]{\mydatasPIP};
              \pgfplotstableread{data3PIP.txt}\mydatatPIP;
              \addplot+[only marks] table [x=uniformity,y=alignment]{\mydatatPIP};
              \pgfplotstableread{data4PIP.txt}\mydatafPIP;
              \addplot+[only marks] table [x=uniformity,y=alignment]{\mydatafPIP};
              \pgfplotstableread{data5PIP.txt}\mydataffPIP;
              \addplot+[only marks] table [x=uniformity,y=alignment]{\mydataffPIP};
              \pgfplotstableread{data6PIP.txt}\mydatasPIP;
              \addplot+[only marks] table [x=uniformity,y=alignment]{\mydatasPIP};
            \end{axis}
          \end{tikzpicture}
     \end{minipage}
     
     \begin{subfigure}{0.3\textwidth}
         \flushleft
          \begin{tikzpicture}
            \begin{axis}[
                title={RLP},
                xlabel=$\ell_{uniform}$,
                ylabel=$\ell_{align}$,
                width=4.2cm,
                height=4.2cm,
                legend style={at={(1.4,0.5)},anchor= west},
                axis equal
              ]
              \pgfplotstableread{data2RLP.txt}\mydatasRLP;
              \addplot+[only marks] table [x=uniformity,y=alignment]{\mydatasRLP};
              \addlegendentry{$\rm UBM$}
              \pgfplotstableread{data3RLP.txt}\mydatatRLP;
              \addplot+[only marks] table [x=uniformity,y=alignment]{\mydatatRLP};
              \addlegendentry{$\rm BERT_{mini}$ }
              \pgfplotstableread{data4RLP.txt}\mydatafRLP;
              \addplot+[only marks] table [x=uniformity,y=alignment]{\mydatafRLP};
              \addlegendentry{$\rm BERT_{8layer}$ }
              \pgfplotstableread{data5RLP.txt}\mydataffRLP;
              \addplot+[only marks] table [x=uniformity,y=alignment]{\mydataffRLP};
              \addlegendentry{$\rm BERT_{mini\_domain}$ }
              \pgfplotstableread{data6RLP.txt}\mydatasRLP;
              \addplot+[only marks] table [x=uniformity,y=alignment]{\mydatasRLP};
              \addlegendentry{$\rm BERT_{8layer\_domain} $}
            \end{axis}
          \end{tikzpicture}
     \end{subfigure}
     \vspace{-0.2cm}
        \caption{Comparison of $\ell_{uniform}$-$\ell_{align}$ pairs on the three tasks' test sets. For both values, the smaller the better.}
        \label{fig:uni-ali}
\end{figure}
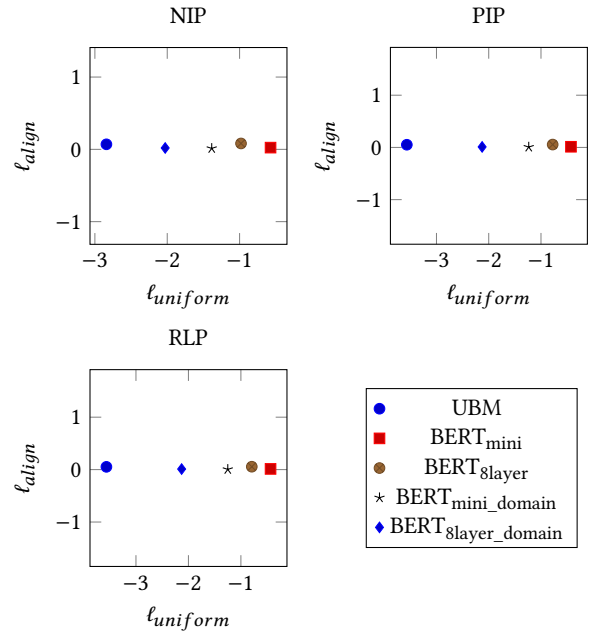

\citet{uniform_align} identified two key properties of contrastive loss: alignment and uniformity, which are used by different researchers as a tool to understand contrastive learning. They provided mathematical proof and experiment results showing that smaller alignment loss and smaller uniformity loss correlate with smaller contrastive loss. With alignment loss, the encoder is trained towards putting positive pairs closer. While with uniformity loss, the encoder is trained towards feature distribution that preserves maximal information. Both are important properties for representation learning. To understand further on our method using the two metrics, we prepare Figure \ref{fig:uni-ali}. We adapt the below definitions from \cite{uniform_align} and use them for the calculation, where $x$,$z$ refers to a data pair from the data samples. 
\begin{equation*}
  \ell_{\rm align}\triangleq \mathop{\mathbb{E}}\limits_{(x,x^+)\sim p_{\rm pos}}\|f(x)-f(x^+)\|^2.
\end{equation*}
\begin{equation*}
  \ell_{\rm uniform}\triangleq \log\mathop{\mathbb{E}}\limits_{x,z\mathop{\sim}\limits^{i.i.d} p_{\rm data}}e^{-2\|f(x)-f(z)\|^2}.
\end{equation*}

We calculated the alignment and uniformity values for UBM, ${BERT_{mini}}$, ${BERT_{8layer}}$, $BERT_{mini\_domain}$ and ${BERT_{8layer\_domain}}$. The data samples are each task's test data augmented with \textit{Action and Item Token Masking}. We also run with \textit{Behavior Reordering} augmentation method and the results show similar trend, hence it is omitted.

As it can be seen from Figure \ref{fig:uni-ali}, their alignment losses are quite similar for different models, while their uniformity losses differentiate them. UBM is with the smallest uniformity loss, which lead to the best overall performance. Domain adapted models have smaller uniformity losses also, although it is not obviously reflected in the NIP performance results for some metrics.

\vspace{-0.35cm}

\vspace{-0.4cm}

%% file: main.bbl

\begin{thebibliography}{44}


\ifx \showCODEN    \undefined \def \showCODEN     #1{\unskip}     \fi
\ifx \showDOI      \undefined \def \showDOI       #1{#1}\fi
\ifx \showISBNx    \undefined \def \showISBNx     #1{\unskip}     \fi
\ifx \showISBNxiii \undefined \def \showISBNxiii  #1{\unskip}     \fi
\ifx \showISSN     \undefined \def \showISSN      #1{\unskip}     \fi
\ifx \showLCCN     \undefined \def \showLCCN      #1{\unskip}     \fi
\ifx \shownote     \undefined \def \shownote      #1{#1}          \fi
\ifx \showarticletitle \undefined \def \showarticletitle #1{#1}   \fi
\ifx \showURL      \undefined \def \showURL       {\relax}        \fi
\providecommand\bibfield[2]{#2}
\providecommand\bibinfo[2]{#2}
\providecommand\natexlab[1]{#1}
\providecommand\showeprint[2][]{arXiv:#2}

\bibitem[Bhargava et~al\mbox{.}(2021)]%
        {bhargava2021generalization}
\bibfield{author}{\bibinfo{person}{Prajjwal Bhargava}, \bibinfo{person}{Aleksandr Drozd}, {and} \bibinfo{person}{Anna Rogers}.} \bibinfo{year}{2021}\natexlab{}.
\newblock \showarticletitle{Generalization in {NLI}: Ways (Not) To Go Beyond Simple Heuristics}. In \bibinfo{booktitle}{\emph{Proceedings of the Second Workshop on Insights from Negative Results in NLP}}. \bibinfo{publisher}{ACL}, \bibinfo{pages}{125--135}.
\newblock


\bibitem[Bianchi et~al\mbox{.}(2021)]%
        {prod2bert}
\bibfield{author}{\bibinfo{person}{Federico Bianchi}, \bibinfo{person}{Bingqing Yu}, {and} \bibinfo{person}{Jacopo Tagliabue}.} \bibinfo{year}{2021}\natexlab{}.
\newblock \showarticletitle{{BERT} Goes Shopping: Comparing Distributional Models for Product Representations}. In \bibinfo{booktitle}{\emph{Proceedings of The 4th Workshop on e-Commerce and NLP}}. \bibinfo{publisher}{ACL}.
\newblock


\bibitem[Brown et~al\mbox{.}(2020)]%
        {gpt-3}
\bibfield{author}{\bibinfo{person}{Tom Brown}, \bibinfo{person}{Benjamin Mann}, \bibinfo{person}{Nick Ryder}, \bibinfo{person}{Melanie Subbiah}, \bibinfo{person}{Jared~D Kaplan}, \bibinfo{person}{Prafulla Dhariwal}, \bibinfo{person}{Arvind Neelakantan}, \bibinfo{person}{Pranav Shyam}, \bibinfo{person}{Girish Sastry}, \bibinfo{person}{Amanda Askell}, \bibinfo{person}{Sandhini Agarwal}, \bibinfo{person}{Ariel Herbert-Voss}, \bibinfo{person}{Gretchen Krueger}, \bibinfo{person}{Tom Henighan}, \bibinfo{person}{Rewon Child}, \bibinfo{person}{Aditya Ramesh}, \bibinfo{person}{Daniel Ziegler}, \bibinfo{person}{Jeffrey Wu}, \bibinfo{person}{Clemens Winter}, \bibinfo{person}{Chris Hesse}, \bibinfo{person}{Mark Chen}, \bibinfo{person}{Eric Sigler}, \bibinfo{person}{Mateusz Litwin}, \bibinfo{person}{Scott Gray}, \bibinfo{person}{Benjamin Chess}, \bibinfo{person}{Jack Clark}, \bibinfo{person}{Christopher Berner}, \bibinfo{person}{Sam McCandlish}, \bibinfo{person}{Alec Radford}, \bibinfo{person}{Ilya Sutskever}, {and}
  \bibinfo{person}{Dario Amodei}.} \bibinfo{year}{2020}\natexlab{}.
\newblock \showarticletitle{Language Models are Few-Shot Learners}. In \bibinfo{booktitle}{\emph{Advances in Neural Information Processing Systems}}. \bibinfo{pages}{1877--1901}.
\newblock


\bibitem[Chen et~al\mbox{.}(2020)]%
        {SimCLR}
\bibfield{author}{\bibinfo{person}{Ting Chen}, \bibinfo{person}{Simon Kornblith}, \bibinfo{person}{Mohammad Norouzi}, {and} \bibinfo{person}{Geoffrey Hinton}.} \bibinfo{year}{2020}\natexlab{}.
\newblock \showarticletitle{A Simple Framework for Contrastive Learning of Visual Representations}. In \bibinfo{booktitle}{\emph{Proceedings of the 37th International Conference on Machine Learning}}. \bibinfo{publisher}{PMLR}, \bibinfo{pages}{1597--1607}.
\newblock


\bibitem[Chen et~al\mbox{.}(2019)]%
        {AIR}
\bibfield{author}{\bibinfo{person}{Tong Chen}, \bibinfo{person}{Hongzhi Yin}, \bibinfo{person}{Hongxu Chen}, \bibinfo{person}{Rui Yan}, \bibinfo{person}{Quoc Viet~Hung Nguyen}, {and} \bibinfo{person}{Xue Li}.} \bibinfo{year}{2019}\natexlab{}.
\newblock \showarticletitle{AIR: Attentional Intention-Aware Recommender Systems}. In \bibinfo{booktitle}{\emph{2019 IEEE 35th International Conference on Data Engineering}}. \bibinfo{pages}{304--315}.
\newblock


\bibitem[Clark et~al\mbox{.}(2020)]%
        {clark2020electra}
\bibfield{author}{\bibinfo{person}{Kevin Clark}, \bibinfo{person}{Minh-Thang Luong}, \bibinfo{person}{Quoc~V. Le}, {and} \bibinfo{person}{Christopher~D. Manning}.} \bibinfo{year}{2020}\natexlab{}.
\newblock \showarticletitle{{ELECTRA}: Pre-training Text Encoders as Discriminators Rather Than Generators}. In \bibinfo{booktitle}{\emph{Proceedings of the 2020 International Conference on Learning Representations}}.
\newblock


\bibitem[Devlin et~al\mbox{.}(2019)]%
        {bert}
\bibfield{author}{\bibinfo{person}{Jacob Devlin}, \bibinfo{person}{Ming-Wei Chang}, \bibinfo{person}{Kenton Lee}, {and} \bibinfo{person}{Kristina Toutanova}.} \bibinfo{year}{2019}\natexlab{}.
\newblock \showarticletitle{BERT: Pre-training of Deep Bidirectional Transformers for Language Understanding}.
\newblock \bibinfo{journal}{\emph{ArXiv}} (\bibinfo{year}{2019}).
\newblock


\bibitem[Diamantaras et~al\mbox{.}(2021)]%
        {Diamantaras}
\bibfield{author}{\bibinfo{person}{Kostas Diamantaras}, \bibinfo{person}{Mike Salampasis}, \bibinfo{person}{Alkiviadis Katsalis}, {and} \bibinfo{person}{Konstantinos Christantonis}.} \bibinfo{year}{2021}\natexlab{}.
\newblock \showarticletitle{Predicting Shopping Intent of e-Commerce Users using LSTM Recurrent Neural Networks}. In \bibinfo{booktitle}{\emph{10th International Conference on Data Science, Technology and Applications}}.
\newblock


\bibitem[Dong et~al\mbox{.}(2022)]%
        {dong2022table}
\bibfield{author}{\bibinfo{person}{Haoyu Dong}, \bibinfo{person}{Zhoujun Cheng}, \bibinfo{person}{Xinyi He}, \bibinfo{person}{Mengyu Zhou}, \bibinfo{person}{Anda Zhou}, \bibinfo{person}{Fan Zhou}, \bibinfo{person}{Ao Liu}, \bibinfo{person}{Shi Han}, {and} \bibinfo{person}{Dongmei Zhang}.} \bibinfo{year}{2022}\natexlab{}.
\newblock \showarticletitle{Table Pre-training: A Survey on Model Architectures, Pre-training Objectives, and Downstream Tasks}. In \bibinfo{booktitle}{\emph{IJCAI'2022 SURVEY TRACK}}.
\newblock


\bibitem[Du et~al\mbox{.}(2022)]%
        {pre-train_survey}
\bibfield{author}{\bibinfo{person}{Yifan Du}, \bibinfo{person}{Zikang Liu}, \bibinfo{person}{Junyi Li}, {and} \bibinfo{person}{Wayne~Xin Zhao}.} \bibinfo{year}{2022}\natexlab{}.
\newblock \bibinfo{title}{A Survey of Vision-Language Pre-Trained Models}.
\newblock
\newblock


\bibitem[Erhan et~al\mbox{.}(2010)]%
        {Dumitru_why}
\bibfield{author}{\bibinfo{person}{Dumitru Erhan}, \bibinfo{person}{Yoshua Bengio}, \bibinfo{person}{Aaron Courville}, \bibinfo{person}{Pierre-Antoine Manzagol}, \bibinfo{person}{Pascal Vincent}, {and} \bibinfo{person}{Samy Bengio}.} \bibinfo{year}{2010}\natexlab{}.
\newblock \showarticletitle{Why Does Unsupervised Pre-training Help Deep Learning?}
\newblock \bibinfo{journal}{\emph{Journal of Machine Learning Research}} (\bibinfo{year}{2010}), \bibinfo{pages}{625--660}.
\newblock


\bibitem[Gao et~al\mbox{.}(2021)]%
        {simcse}
\bibfield{author}{\bibinfo{person}{Tianyu Gao}, \bibinfo{person}{Xingcheng Yao}, {and} \bibinfo{person}{Danqi Chen}.} \bibinfo{year}{2021}\natexlab{}.
\newblock \showarticletitle{{S}im{CSE}: Simple Contrastive Learning of Sentence Embeddings}. In \bibinfo{booktitle}{\emph{Proceedings of the 2021 Conference on Empirical Methods in Natural Language Processing}}. \bibinfo{publisher}{ACL}, \bibinfo{pages}{6894--6910}.
\newblock


\bibitem[Gupta et~al\mbox{.}(2019)]%
        {niser}
\bibfield{author}{\bibinfo{person}{Priyanka Gupta}, \bibinfo{person}{Diksha Garg}, \bibinfo{person}{Pankaj Malhotra}, \bibinfo{person}{Lovekesh Vig}, {and} \bibinfo{person}{Gautam Shroff}.} \bibinfo{year}{2019}\natexlab{}.
\newblock \showarticletitle{{NISER:} Normalized Item and Session Representations with Graph Neural Networks}.
\newblock \bibinfo{journal}{\emph{CoRR}} (\bibinfo{year}{2019}).
\newblock


\bibitem[Gupta and Maji(2020)]%
        {Gupta2020RLP}
\bibfield{author}{\bibinfo{person}{Shashank Gupta} {and} \bibinfo{person}{Subhadeep Maji}.} \bibinfo{year}{2020}\natexlab{}.
\newblock \showarticletitle{Predicting Session Length for Product Search on E-Commerce Platform}. In \bibinfo{booktitle}{\emph{Proceedings of the 43rd International ACM SIGIR Conference on Research and Development in Information Retrieval}}. \bibinfo{publisher}{ACM}.
\newblock


\bibitem[Hadsell et~al\mbox{.}(2006)]%
        {hadsell2006dimensionality}
\bibfield{author}{\bibinfo{person}{Raia Hadsell}, \bibinfo{person}{Sumit Chopra}, {and} \bibinfo{person}{Yann LeCun}.} \bibinfo{year}{2006}\natexlab{}.
\newblock \showarticletitle{Dimensionality reduction by learning an invariant mapping}. In \bibinfo{booktitle}{\emph{CVPR}}, Vol.~\bibinfo{volume}{2}. \bibinfo{pages}{1735--1742}.
\newblock


\bibitem[Hatt and Feuerriegel(2020)]%
        {m3pp}
\bibfield{author}{\bibinfo{person}{Tobias Hatt} {and} \bibinfo{person}{Stefan Feuerriegel}.} \bibinfo{year}{2020}\natexlab{}.
\newblock \showarticletitle{Early Detection of User Exits from Clickstream Data: A Markov Modulated Marked Point Process Model}. In \bibinfo{booktitle}{\emph{Proceedings of The Web Conference 2020}}. \bibinfo{publisher}{ACM}, \bibinfo{pages}{1671–1681}.
\newblock


\bibitem[He et~al\mbox{.}(2020)]%
        {he2020momentum}
\bibfield{author}{\bibinfo{person}{Kaiming He}, \bibinfo{person}{Haoqi Fan}, \bibinfo{person}{Yuxin Wu}, \bibinfo{person}{Saining Xie}, {and} \bibinfo{person}{Ross Girshick}.} \bibinfo{year}{2020}\natexlab{}.
\newblock \showarticletitle{Momentum contrast for unsupervised visual representation learning}. In \bibinfo{booktitle}{\emph{CVPR}}. \bibinfo{pages}{9729--9738}.
\newblock


\bibitem[He et~al\mbox{.}(2015)]%
        {resnet}
\bibfield{author}{\bibinfo{person}{Kaiming He}, \bibinfo{person}{Xiangyu Zhang}, \bibinfo{person}{Shaoqing Ren}, {and} \bibinfo{person}{Jian Sun}.} \bibinfo{year}{2015}\natexlab{}.
\newblock \bibinfo{title}{Deep Residual Learning for Image Recognition}.
\newblock
\newblock


\bibitem[Hidasi et~al\mbox{.}(2016)]%
        {Hidasi2016SessionbasedRW}
\bibfield{author}{\bibinfo{person}{Bal{\'a}zs Hidasi}, \bibinfo{person}{Alexandros Karatzoglou}, \bibinfo{person}{Linas Baltrunas}, {and} \bibinfo{person}{Domonkos Tikk}.} \bibinfo{year}{2016}\natexlab{}.
\newblock \showarticletitle{Session-based Recommendations with Recurrent Neural Networks}.
\newblock \bibinfo{journal}{\emph{CoRR}} (\bibinfo{year}{2016}).
\newblock


\bibitem[Jaiswal et~al\mbox{.}(2021)]%
        {technologies9010002}
\bibfield{author}{\bibinfo{person}{Ashish Jaiswal}, \bibinfo{person}{Ashwin~Ramesh Babu}, \bibinfo{person}{Mohammad~Zaki Zadeh}, \bibinfo{person}{Debapriya Banerjee}, {and} \bibinfo{person}{Fillia Makedon}.} \bibinfo{year}{2021}\natexlab{}.
\newblock \showarticletitle{A Survey on Contrastive Self-Supervised Learning}.
\newblock \bibinfo{journal}{\emph{Technologies}} (\bibinfo{year}{2021}).
\newblock


\bibitem[Kalantidis et~al\mbox{.}(2020)]%
        {kalantidis2020hard}
\bibfield{author}{\bibinfo{person}{Yannis Kalantidis}, \bibinfo{person}{Mert~Bulent Sariyildiz}, \bibinfo{person}{Noe Pion}, \bibinfo{person}{Philippe Weinzaepfel}, {and} \bibinfo{person}{Diane Larlus}.} \bibinfo{year}{2020}\natexlab{}.
\newblock \showarticletitle{Hard Negative Mixing for Contrastive Learning}. In \bibinfo{booktitle}{\emph{Neural Information Processing Systems}}.
\newblock


\bibitem[Lan et~al\mbox{.}(2019)]%
        {lan2019albert}
\bibfield{author}{\bibinfo{person}{Zhenzhong Lan}, \bibinfo{person}{Mingda Chen}, \bibinfo{person}{Sebastian Goodman}, \bibinfo{person}{Kevin Gimpel}, \bibinfo{person}{Piyush Sharma}, {and} \bibinfo{person}{Radu Soricut}.} \bibinfo{year}{2019}\natexlab{}.
\newblock \showarticletitle{ALBERT: A Lite BERT for Self-supervised Learning of Language Representations}. In \bibinfo{booktitle}{\emph{International Conference on Learning Representations}}.
\newblock


\bibitem[Li et~al\mbox{.}(2022)]%
        {lta}
\bibfield{author}{\bibinfo{person}{Zixuan Li}, \bibinfo{person}{Lizi Liao}, {and} \bibinfo{person}{Tat-Seng Chua}.} \bibinfo{year}{2022}\natexlab{}.
\newblock \showarticletitle{Learning to Ask Critical Questions for Assisting Product Search}. In \bibinfo{booktitle}{\emph{Proceedings of ACM SIGIR Workshop on eCommerce (SIGIR eCom’22)}}.
\newblock


\bibitem[Liu et~al\mbox{.}(2018)]%
        {stamp}
\bibfield{author}{\bibinfo{person}{Qiao Liu}, \bibinfo{person}{Yifu Zeng}, \bibinfo{person}{Refuoe Mokhosi}, {and} \bibinfo{person}{Haibin Zhang}.} \bibinfo{year}{2018}\natexlab{}.
\newblock \showarticletitle{STAMP: Short-Term Attention/Memory Priority Model for Session-Based Recommendation}. In \bibinfo{booktitle}{\emph{Proceedings of the 24th ACM SIGKDD International Conference on Knowledge Discovery and Data Mining}}. \bibinfo{publisher}{ACM}, \bibinfo{pages}{1831–1839}.
\newblock


\bibitem[Liu et~al\mbox{.}(2019)]%
        {roberta}
\bibfield{author}{\bibinfo{person}{Yinhan Liu}, \bibinfo{person}{Myle Ott}, \bibinfo{person}{Naman Goyal}, \bibinfo{person}{Jingfei Du}, \bibinfo{person}{Mandar Joshi}, \bibinfo{person}{Danqi Chen}, \bibinfo{person}{Omer Levy}, \bibinfo{person}{Mike Lewis}, \bibinfo{person}{Luke Zettlemoyer}, {and} \bibinfo{person}{Veselin Stoyanov}.} \bibinfo{year}{2019}\natexlab{}.
\newblock \bibinfo{title}{RoBERTa: A Robustly Optimized BERT Pretraining Approach}.
\newblock
\newblock


\bibitem[Peters et~al\mbox{.}(2018)]%
        {elmo}
\bibfield{author}{\bibinfo{person}{Matthew~E. Peters}, \bibinfo{person}{Mark Neumann}, \bibinfo{person}{Mohit Iyyer}, \bibinfo{person}{Matt Gardner}, \bibinfo{person}{Christopher Clark}, \bibinfo{person}{Kenton Lee}, {and} \bibinfo{person}{Luke Zettlemoyer}.} \bibinfo{year}{2018}\natexlab{}.
\newblock \showarticletitle{Deep Contextualized Word Representations}. In \bibinfo{booktitle}{\emph{Proceedings of the 2018 Conference of the North {A}merican Chapter of the Association for Computational Linguistics: Human Language Technologies}}. \bibinfo{publisher}{ACL}, \bibinfo{pages}{2227--2237}.
\newblock


\bibitem[Rendle et~al\mbox{.}(2010)]%
        {rendle2010fpmc}
\bibfield{author}{\bibinfo{person}{Steffen Rendle}, \bibinfo{person}{Christoph Freudenthaler}, {and} \bibinfo{person}{Lars Schmidt-Thieme}.} \bibinfo{year}{2010}\natexlab{}.
\newblock \showarticletitle{Factorizing Personalized Markov Chains for Next-Basket Recommendation}. In \bibinfo{booktitle}{\emph{Proceedings of the 19th International Conference on World Wide Web}}. \bibinfo{publisher}{ACM}, \bibinfo{pages}{811–820}.
\newblock


\bibitem[Robinson et~al\mbox{.}(2021)]%
        {robinson2021contrastive}
\bibfield{author}{\bibinfo{person}{Joshua~David Robinson}, \bibinfo{person}{Ching-Yao Chuang}, \bibinfo{person}{Suvrit Sra}, {and} \bibinfo{person}{Stefanie Jegelka}.} \bibinfo{year}{2021}\natexlab{}.
\newblock \showarticletitle{Contrastive Learning with Hard Negative Samples}. In \bibinfo{booktitle}{\emph{International Conference on Learning Representations}}.
\newblock


\bibitem[Sakar et~al\mbox{.}(2019)]%
        {SakarPIP}
\bibfield{author}{\bibinfo{person}{C.~Okan Sakar}, \bibinfo{person}{S. Polat}, \bibinfo{person}{Mete Katircioglu}, {and} \bibinfo{person}{Yomi Kastro}.} \bibinfo{year}{2019}\natexlab{}.
\newblock \showarticletitle{Real-time prediction of online shoppers’ purchasing intention using multilayer perceptron and LSTM recurrent neural networks}.
\newblock \bibinfo{journal}{\emph{Neural Computing and Applications}} (\bibinfo{year}{2019}).
\newblock


\bibitem[Shani et~al\mbox{.}(2002)]%
        {mdp2002}
\bibfield{author}{\bibinfo{person}{Guy Shani}, \bibinfo{person}{Ronen~I. Brafman}, {and} \bibinfo{person}{David Heckerman}.} \bibinfo{year}{2002}\natexlab{}.
\newblock \showarticletitle{An MDP-Based Recommender System}. In \bibinfo{booktitle}{\emph{Proceedings of the 18th Conference on Uncertainty in Artificial Intelligence}}. \bibinfo{publisher}{Morgan Kaufmann Publishers Inc.}
\newblock


\bibitem[Simonyan and Zisserman(2015)]%
        {SimonyanZ14a}
\bibfield{author}{\bibinfo{person}{Karen Simonyan} {and} \bibinfo{person}{Andrew Zisserman}.} \bibinfo{year}{2015}\natexlab{}.
\newblock \showarticletitle{Very Deep Convolutional Networks for Large-Scale Image Recognition}. In \bibinfo{booktitle}{\emph{3rd International Conference on Learning Representations}}.
\newblock


\bibitem[Sun et~al\mbox{.}(2019)]%
        {bert4rec}
\bibfield{author}{\bibinfo{person}{Fei Sun}, \bibinfo{person}{Jun Liu}, \bibinfo{person}{Jian Wu}, \bibinfo{person}{Changhua Pei}, \bibinfo{person}{Xiao Lin}, \bibinfo{person}{Wenwu Ou}, {and} \bibinfo{person}{Peng Jiang}.} \bibinfo{year}{2019}\natexlab{}.
\newblock \showarticletitle{BERT4Rec: Sequential Recommendation with Bidirectional Encoder Representations from Transformer}. In \bibinfo{booktitle}{\emph{Proceedings of the 28th ACM International Conference on Information and Knowledge Management}}. \bibinfo{publisher}{ACM}, \bibinfo{pages}{1441–1450}.
\newblock


\bibitem[Tan and Bansal(2019)]%
        {lxmert}
\bibfield{author}{\bibinfo{person}{Hao Tan} {and} \bibinfo{person}{Mohit Bansal}.} \bibinfo{year}{2019}\natexlab{}.
\newblock \showarticletitle{{LXMERT}: Learning Cross-Modality Encoder Representations from Transformers}. In \bibinfo{booktitle}{\emph{Proceedings of the 2019 Conference on Empirical Methods in Natural Language Processing and the 9th International Joint Conference on Natural Language Processing}}. \bibinfo{publisher}{ACL}, \bibinfo{pages}{5100--5111}.
\newblock


\bibitem[Tang and Wang(2018)]%
        {CNN2018}
\bibfield{author}{\bibinfo{person}{Jiaxi Tang} {and} \bibinfo{person}{Ke Wang}.} \bibinfo{year}{2018}\natexlab{}.
\newblock \showarticletitle{Personalized Top-N Sequential Recommendation via Convolutional Sequence Embedding}. In \bibinfo{booktitle}{\emph{Proceedings of the Eleventh ACM International Conference on Web Search and Data Mining}}. \bibinfo{publisher}{ACM}, \bibinfo{pages}{565–573}.
\newblock


\bibitem[Tian et~al\mbox{.}(2020)]%
        {goodview}
\bibfield{author}{\bibinfo{person}{Yonglong Tian}, \bibinfo{person}{Chen Sun}, \bibinfo{person}{Ben Poole}, \bibinfo{person}{Dilip Krishnan}, \bibinfo{person}{Cordelia Schmid}, {and} \bibinfo{person}{Phillip Isola}.} \bibinfo{year}{2020}\natexlab{}.
\newblock \showarticletitle{What Makes for Good Views for Contrastive Learning?}. In \bibinfo{booktitle}{\emph{Proceedings of the 34th International Conference on Neural Information Processing Systems}}. \bibinfo{publisher}{Curran Associates Inc.}
\newblock


\bibitem[Turc et~al\mbox{.}(2019)]%
        {DBLP:journals/corr/abs-1908-08962}
\bibfield{author}{\bibinfo{person}{Iulia Turc}, \bibinfo{person}{Ming{-}Wei Chang}, \bibinfo{person}{Kenton Lee}, {and} \bibinfo{person}{Kristina Toutanova}.} \bibinfo{year}{2019}\natexlab{}.
\newblock \showarticletitle{Well-Read Students Learn Better: The Impact of Student Initialization on Knowledge Distillation}.
\newblock \bibinfo{journal}{\emph{CoRR}} (\bibinfo{year}{2019}).
\newblock


\bibitem[Wang et~al\mbox{.}(2021)]%
        {srs_survey}
\bibfield{author}{\bibinfo{person}{Shoujin Wang}, \bibinfo{person}{Longbing Cao}, \bibinfo{person}{Yan Wang}, \bibinfo{person}{Quan~Z. Sheng}, \bibinfo{person}{Mehmet~A. Orgun}, {and} \bibinfo{person}{Defu Lian}.} \bibinfo{year}{2021}\natexlab{}.
\newblock \showarticletitle{A Survey on Session-Based Recommender Systems}.
\newblock \bibinfo{journal}{\emph{Comput. Surveys}} (\bibinfo{year}{2021}).
\newblock


\bibitem[Wang et~al\mbox{.}(2019)]%
        {ijcai2019p883}
\bibfield{author}{\bibinfo{person}{Shoujin Wang}, \bibinfo{person}{Liang Hu}, \bibinfo{person}{Yan Wang}, \bibinfo{person}{Longbing Cao}, \bibinfo{person}{Quan~Z. Sheng}, {and} \bibinfo{person}{Mehmet Orgun}.} \bibinfo{year}{2019}\natexlab{}.
\newblock \showarticletitle{Sequential Recommender Systems: Challenges, Progress and Prospects}. In \bibinfo{booktitle}{\emph{Proceedings of the Twenty-Eighth International Joint Conference on Artificial Intelligence, {IJCAI-19}}}. \bibinfo{publisher}{IJCAI}, \bibinfo{pages}{6332--6338}.
\newblock


\bibitem[Wang and Isola(2020)]%
        {uniform_align}
\bibfield{author}{\bibinfo{person}{Tongzhou Wang} {and} \bibinfo{person}{Phillip Isola}.} \bibinfo{year}{2020}\natexlab{}.
\newblock \showarticletitle{Understanding Contrastive Representation Learning through Alignment and Uniformity on the Hypersphere}. In \bibinfo{booktitle}{\emph{Proceedings of the 37th International Conference on Machine Learning}}. \bibinfo{publisher}{JMLR}.
\newblock


\bibitem[Wu et~al\mbox{.}(2019)]%
        {Wu_Tang_Zhu_Wang_Xie_Tan_2019}
\bibfield{author}{\bibinfo{person}{Shu Wu}, \bibinfo{person}{Yuyuan Tang}, \bibinfo{person}{Yanqiao Zhu}, \bibinfo{person}{Liang Wang}, \bibinfo{person}{Xing Xie}, {and} \bibinfo{person}{Tieniu Tan}.} \bibinfo{year}{2019}\natexlab{}.
\newblock \showarticletitle{Session-Based Recommendation with Graph Neural Networks}.
\newblock \bibinfo{journal}{\emph{Proceedings of the AAAI Conference on Artificial Intelligence}} (\bibinfo{year}{2019}), \bibinfo{pages}{346--353}.
\newblock


\bibitem[Yan et~al\mbox{.}(2021a)]%
        {yan2021consert}
\bibfield{author}{\bibinfo{person}{Yuanmeng Yan}, \bibinfo{person}{Rumei Li}, \bibinfo{person}{Sirui Wang}, \bibinfo{person}{Fuzheng Zhang}, \bibinfo{person}{Wei Wu}, {and} \bibinfo{person}{Weiran Xu}.} \bibinfo{year}{2021}\natexlab{a}.
\newblock \showarticletitle{ConSERT: A Contrastive Framework for Self-Supervised Sentence Representation Transfer}. In \bibinfo{booktitle}{\emph{ACL}}. \bibinfo{pages}{5065--5075}.
\newblock


\bibitem[Yan et~al\mbox{.}(2021b)]%
        {yan-etal-2021-consert}
\bibfield{author}{\bibinfo{person}{Yuanmeng Yan}, \bibinfo{person}{Rumei Li}, \bibinfo{person}{Sirui Wang}, \bibinfo{person}{Fuzheng Zhang}, \bibinfo{person}{Wei Wu}, {and} \bibinfo{person}{Weiran Xu}.} \bibinfo{year}{2021}\natexlab{b}.
\newblock \showarticletitle{{C}on{SERT}: A Contrastive Framework for Self-Supervised Sentence Representation Transfer}. In \bibinfo{booktitle}{\emph{Proceedings of the 59th Annual Meeting of the Association for Computational Linguistics and the 11th International Joint Conference on Natural Language Processing}}. \bibinfo{publisher}{ACL}, \bibinfo{pages}{5065--5075}.
\newblock


\bibitem[Yao et~al\mbox{.}(2019)]%
        {yao2019kg}
\bibfield{author}{\bibinfo{person}{Liang Yao}, \bibinfo{person}{Chengsheng Mao}, {and} \bibinfo{person}{Yuan Luo}.} \bibinfo{year}{2019}\natexlab{}.
\newblock \showarticletitle{KG-BERT: BERT for knowledge graph completion}.
\newblock \bibinfo{journal}{\emph{arXiv}} (\bibinfo{year}{2019}).
\newblock


\bibitem[Zhu et~al\mbox{.}(2021)]%
        {Zhu2021ContrastiveLO}
\bibfield{author}{\bibinfo{person}{Yutao Zhu}, \bibinfo{person}{Jianyun Nie}, \bibinfo{person}{Zhicheng Dou}, \bibinfo{person}{Zhengyi Ma}, \bibinfo{person}{Xinyu Zhang}, \bibinfo{person}{Pan Du}, \bibinfo{person}{Xiaochen Zuo}, {and} \bibinfo{person}{Hao Jiang}.} \bibinfo{year}{2021}\natexlab{}.
\newblock \showarticletitle{Contrastive Learning of User Behavior Sequence for Context-Aware Document Ranking}.
\newblock \bibinfo{journal}{\emph{Proceedings of the 30th ACM International Conference on Information and Knowledge Management}} (\bibinfo{year}{2021}).
\newblock


\end{thebibliography}
